\documentclass[sigconf]{acmart}
\AtBeginDocument{%
  }

\copyrightyear{2026}
\acmYear{2026}
\setcopyright{cc}
\setcctype{by}
\acmConference[UIST '26]{The 39th Annual ACM Symposium on User Interface Software and Technology}{November 02--05, 2026}{Detroit, MI, USA}
\acmBooktitle{The 39th Annual ACM Symposium on User Interface Software and Technology (UIST '26), November 02--05, 2026, Detroit, MI, USA}
\acmDOI{10.1145/3830398.3830611}
\acmISBN{979-8-4007-2856-3/2026/11}

\usepackage{subcaption}
\usepackage{listings}
\usepackage{url}
\begin{document}

\title{IRIS: Navigating and Reflecting on Writing Traces Using Intelligent Document Histories}

\author{David Zhou}
\affiliation{%
 \institution{University of Illinois Urbana-Champaign}
 \city{Urbana}
 \state{IL}
 \country{USA}
 }
 \email{david23@illinois.edu}

\author{Andrew Chen}
\affiliation{%
 \institution{University of Illinois Urbana-Champaign}
 \city{Urbana}
 \state{IL}
 \country{USA}
 }
 \email{andrew83@illinois.edu}

\author{John Joon Young Chung}
\affiliation{%
 \institution{Midjourney}
 \city{San Francisco}
 \state{CA}
 \country{USA}
 }
 \email{jchung@midjourney.com}

\author{Sarah Sterman}
\affiliation{%
 \institution{University of Illinois Urbana-Champaign}
 \city{Urbana}
 \state{IL}
 \country{USA}
 }
 \email{ssterman@illinois.edu}

\renewcommand{\shortauthors}{Zhou et al.}
\newcommand{\system}{IRIS}
\newcommand{\highlighting}{\texttt{Highlighting}}
\newcommand{\tagging}{\texttt{Filtering}}
\newcommand{\search}{\texttt{Search}}

\begin{abstract}
Much of the text produced throughout the lifetime of a document is impermanent. 
In this paper, we explore how writing activity traces can be made visible and interactive to help writers navigate their document histories and understand their writing processes. Using the Flower and Hayes cognitive process model of writing, \system{} infers writing process states from keystroke logs and presents them using an AI-enhanced version history. \system{} provides three primary interactions: revision highlighting that shows local process histories in-situ, conceptual filters that constrain the version history by process type or topic, and natural language inquiry that lets writers pose reflective questions about their writing and process. Following a formative and a longitudinal study, we find that writers use the interfaces to locate specific revisions and understand the progression of their writing. They use system outputs as interpretive material, relating them to pre-existing beliefs and confirming, challenging, and deepening their understanding of their writing.

\end{abstract}


\begin{teaserfigure}
\centering
  \includegraphics[width=\textwidth]{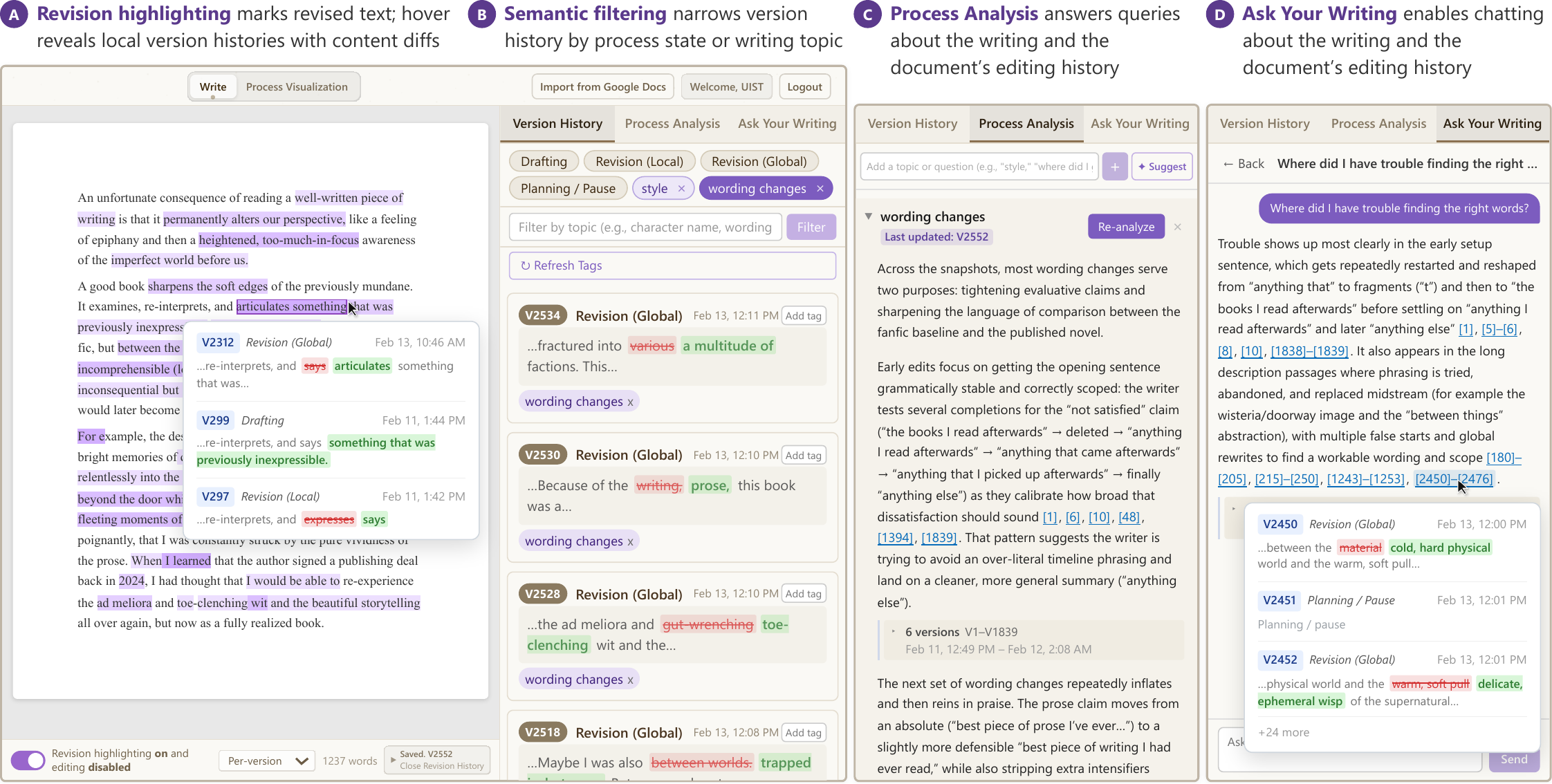}
  \caption{
  \system{} analyzes writing traces to help writers navigate their document histories and reflect on their writing.
  }
  \label{fig:teaser}
\end{teaserfigure}

\newenvironment{myquote}{\begin{quote}\leftskip-14pt\rightskip-14pt}{\end{quote}}
    \newcommand*{\participant}[1]{{\small{\fontfamily{cmss}\selectfont{(#1)}}}}
    \newcommand*{\quoted}[1]{{\small{\fontfamily{cmss}\selectfont{#1}}}}
    \newcommand{\squote}[2]{\begin{myquote}\quoted{#2 \participant{#1}}\end{myquote}}
    \newcommand*{\quotedtext}[1]{\begin{myquote}\small{#1}\end{myquote}}

\newenvironment{myquote2}{}{}
    \newcommand{\sqinline}[2]{\begin{myquote2}\quoted{``#2'' \participant{#1}}\end{myquote2}}
    
\newcommand{\centered}[1]{\begin{tabular}{l} #1 \end{tabular}}

\maketitle

\section{Introduction}
\label{sec:introduction}

The decision making and iteration that go into a final piece of writing cannot be seen in the final output but are instead elements of intermediate writing processes. Word processors sometimes capture this history through features like version histories or track changes, or when writers manually save multiple copies of a file \cite{sterman22towards}. These types of histories are coarse, chronological, and focused on content changes. These limitations constrain what writers can use their histories for: histories are predominantly used as a safety net for making content changes. 

Yet, software tools can capture much richer traces of writing history. From keystrokes to pauses, activity traces can provide evidence of the writing process \cite{leijten13keystroke}.
In this work, we explore how writing activity traces can be used to support the writer as they navigate the history of their documents.  We re-envision interactions with version histories to support both document-history navigation and reflection on one's own writing process. %

Advances in language generation make this type of interaction both possible and urgent.  First, the possibility to analyze fine-grained traces of writing in real time enables the creation of more sophisticated version history interfaces. 
Second, natural language generation tools are leading to a rapid transformation in the roles of writing tools and our relationship to writing. As we see more tools that support particular writing ``best practices'' or automate entire stages of writing \cite{Sudowrite, yuan22wordcraft}, it is timely to expand our approach to writing tools and ask how they can better support diverse writing processes and meaningful engagement with non-text-production tasks, such as writing reflection and history-keeping \cite{reza25cowriting}.%

This paper seeks to answer the question: \textbf{How can we design with writing traces to help writers understand their writing process?}
To address this, we develop a novel AI-powered writing version history tool called \system{}: \underline{I}n-situ \underline{R}evision \underline{I}nquiry \underline{S}ystem. \system{} logs keystroke events, aggregates and categorizes keystrokes into writing process states based upon Flower and Hayes' Cognitive Process Model \cite{flower81cognitive}, and provides these process-informed version states to a retrieval-augmented LLM to enable real-time interaction with history through the lens of writing process theory. %

\system{} explores three primary interactions: 1) \textbf{in-situ} access to the complete writing history; 2) \textbf{filtering} the revision history by author-defined process queries; and 3) \textbf{searching} the revision history by asking questions related to the evolution of the text. While each of these interactions is grounded in existing interaction paradigms, they allow us to explore the novel question of how writers can use process to understand and learn from their writing histories.

We conducted two studies to assess how writers used \system{} to understand their writing process. First, we ran a formative study with $n=14$ participants, in which writers composed a short piece in the tool, then explored the features. This study demonstrated initial uses of the three primary interactions and informed a design revision. Second, we ran a diary study with $n=11$ participants, in which writers composed a longer piece over several writing sessions and used the interface in whichever way they found helpful.  

We show that writers use process-informed version histories as interpretive resources. They relate trace data to pre-existing beliefs about their writing, confirm intuitions about their process, and challenge their personal understanding of their writing in ways that encourage reflection and direct future work. We see that writers' identities and practices shape engagement with the process functionalities, which suggests opportunities for tools that are responsive to the creative process, rather than only to its outputs.

\section{Related Work}
\label{sec:relatedwork}
\system{} captures writing processes from keystroke logs, organizes them within an interactive version history, and visualizes them to support writerly reflection. We review models of the writing process and computational approaches to capturing writing behaviors (Sec.~\ref{sec:rw_capture}), version control systems and the emerging concept of activity traces (Sec. ~\ref{sec:rw_vc}), and writing visualization tools in HCI (Sec.~\ref{sec:rw_vis}).

\subsection{Writing Process and Computers}
\label{sec:rw_capture}

Numerous frameworks characterize writing behaviors and guide pedagogy, from stage-based models \cite{britton75development} to cognitive and post-process theories \cite{berlin87rhetoric, kent93paralogic, kent99post}. 
Among these, process-centric theories such as Flower and Hayes' cognitive process theory \cite{flower81cognitive} have been adopted in HCI to explain how tools influence writing cognition \cite{lee24design, zhang23visar, zhou24aillude} and characterize writers' needs from their tools \cite{gero23social}.
The cognitive process model describes three main processes: \textit{planning}, \textit{translating} (inscribing words onto media), and \textit{reviewing} \cite{flower81cognitive}. 
We use these to structure our system's categorizations of writing traces.
 
Research on capturing writing processes spans think-aloud protocols and writing log analyses. Think-aloud protocols were used by Flower and Hayes \cite{flower80cognition} to study decision-making strategies; these studies partly formed the basis for laboratory keystroke log analysis including InputLog, which records keyboard and mouse actions in Microsoft Word and provides post-hoc analyses such as revision analytics \cite{leijten13keystroke}. We build on keystroke analysis for understanding process; specifically, we adapt the taxonomy of pause and text production states from Zhang et al., who also use the Flower and Hayes model as a theoretical framework to infer cognitive writing process states from keystroke analytics \cite{zhang21using}.
 
Concurrently, keystroke-level traces have been used to study collaborative writing---for instance, how student writers use the document itself to store intermediate material alongside final content \cite{olson17write}. This can be understood through the concept of \emph{text functions}, which describes the multiple roles a text serves during writing, such as ``material'' and ``interim work'' \cite{larsenledet20collaborative}. These efforts show that writing traces are inherently informative, revealing strategies not visible in the finished text.

Recording and analyzing process traces extend beyond typing prose. KiEV \cite{spakov08kiev} tracks gaze to visualize reading behaviors, providing insights on fixation patterns through layered bars denoting reading sequences. CodeProcess charts  visualize code writing by tracking text caret position, offering insight on coding strategies and detecting signs of plagiarism \cite{shrestha22codeprocess}. 
These tools visualize how a task is done and what steps are taken to achieve it.
    
However, these tools are designed primarily to study the processes of \textit{others}---a researcher studying student writers or a teacher observing students. InputLog \cite{leijten13keystroke} and KiEV \cite{spakov08kiev} serve as research instruments; Zhang et al.'s system \cite{zhang21using} classifies student writing quality to investigate time management trends; CodeProcess charts \cite{shrestha22codeprocess} are used by programming instructors. We argue that these insights, if surfaced to writers themselves, are equally beneficial.
Indeed, character insertion visualizations have been shown to increase self-awareness of writing behaviors and collaborative processes \cite{turkay18itero}, suggesting that a process visualization tool designed for writers can serve as a useful metacognitive aid.

\subsection{Creative Process and Version Control}
\label{sec:rw_vc}

Creative work is exploratory and iterative, with later versions building upon earlier ones. On a computer, navigating across states is often implicit---\texttt{undo} and \texttt{redo} navigate between temporally proximal versions. \textit{History management tools} provide explicit state management. For example, Google Docs tracks version states across a document's lifetime through its version history panel.

Version control systems organize changes made to a digital artifact \cite{chacon14git}. While most commonly associated with software code via tools like Git \cite{chacon14git} and Subversion \cite{pilato08subversion}, version control has been applied to diverse creative domains: collaboratively writing books \cite{pe18collaborative}, storing revision histories of images \cite{chen11nonlinear}, 2D vector design \cite{kolari14cambria}, 3D virtual reality scenes \cite{zhang23VRGit}, CAD \cite{chou86unifying}, and story authoring \cite{zund17story}.

Shneiderman argues that rich history-keeping is an important design principle for creativity support tools, noting that intermediate states represent not just steps toward a final artifact but also alternatives that reflect earlier creative decisions \cite{shneiderman07creativity}. Indeed, creative professionals use version histories---slideshows, Adobe Photoshop layers, Git---as materials for exploration and reflection across project lifecycles \cite{sterman22towards}. In Quickpose \cite{rawn23understanding}, Processing 
artists manage version histories as exploratory materials, enabling reflection and experimentation. Reflecting on prior versions ``can reveal changes in the creator's understandings, concepts, or interests'' \cite{sterman22towards}, demonstrating the reflexive nature of creative work.

In writing, Carrera et al. found that watching writing replays facilitated self-reflection, with writers surprised by the effort spent on word-level choices and revision patterns \cite{carrera22watch}. In AI co-writing, ABScribe \cite{reza24ABScribe} attaches local version histories to AI-generated text in-situ as dropdowns, enabling rapid exploration of granular variations. Similarly, DraftMarks uses skeuomorphic metaphors---eraser crumbs for revision, masking tape for AI-generated content---to embed process signals directly on AI co-authored text \cite{siddiqui26draftmarks}.
 
The concept of traces in mediated activity has been theoretically developed by Avdić et al., who examined how records of past activity can serve as mechanisms for collaboration and reflection \cite{avdic21traces}. Hammad et al. later formalized \textit{creative activity traces} as records of creator data produced over the course of a creative process, including artifact iterations, annotations, and reference materials \cite{hammad26tracing}. Related computational approaches include Fuzzy Linkography, which automatically generates design-inspired linkographs from LLM interaction logs to summarize activity traces \cite{smith25fuzzy}. %

These forms of version control show that preserving intermediate states is useful not only for reverting actions, but also for creative exploration and reflection. In writing, however, most systems still treat history as a sequence of snapshots detached from the writing process. This paper seeks to combine version histories with writing process signals to help writers understand their own processes and to navigate past versions using their understanding of their writing process.

\subsection{Writing Visualization in HCI}
\label{sec:rw_vis}
Many HCI writing tools incorporate visualizations to help authors analyze their text. Portrayal \cite{hoque23portrayal} visualizes story characters across narrative time, helping writers track character appearances and interactions. DramatVis Personae \cite{hoque22dramatvis} surfaces potential social biases by mapping descriptive attributes of characters, allowing authors to spot stereotypical patterns as they write. These systems provide live visual feedback, helping authors maintain consistency and awareness in their narratives.
 
Other tools visualize aspects of the writing process to support reflection and revision. DocuViz \cite{wang15docuviz} displays revision history as a color-coded timeline showing authorial contribution, timing, and location. Itero \cite{turkay18itero} visualizes revision statistics to improve self-awareness of editing patterns. In argumentative writing, AL \cite{wambsganss20al} provides interactive views of sentence relationships and persuasiveness scores, while ArgRewrite \cite{zhang16argrewrite} tracks and assesses sentence-level changes across revisions. More recently, Friction visualizes writing feedback as heatmaps beside the text \cite{zhang25friction}, and Texterial reimagines text as a material that users can ``grow, sculpt, and transform'' through LLM-mediated material metaphors \cite{shen26texterial}.
 
Our work builds on these efforts by making visible the cognitive processes of writing. We ground our approach in keystroke writing log analysis \cite{leijten13keystroke} and adopt the Flower and Hayes cognitive model \cite{flower81cognitive} to categorize writing activities. Recent studies of AI-assisted writing have also leveraged this model; Lee et al. \cite{lee24design} map features of writing assistants to cognitive processes, and Gero et al. \cite{gero23social} link writers' unmet needs to stages of the writing process. %
We explore the direct operationalization of the theoretical cognitive model from keystroke logs \cite{zhang21using} and observe how writers interpret this algorithmic representation.

\section{Formative Study}
\label{sec:initialstudy}
We conducted a formative study with fourteen creative writers using an initial system prototype.
To understand how writers make sense of their writing process, we designed and implemented a live writing environment that includes three ways to navigate the version history (Fig. \ref{fig:design_changes}, top). In this section, we describe our design goals and prototype, followed by the study methods and findings. The full system implementation is described in Section \ref{sec:IRIS}.

\subsection{Prototype}

A significant portion of typed text is temporary: once text is revised, writers cannot easily see where effort accumulated and how ideas evolved. To study how writers make sense of their writing process, we built a text editor that infers writing process states from keystrokes and presents them in an interactive version history. 
Process data can be informative, providing a “palette of materials” \cite{sterman22towards} such as previously considered alternatives. The prototype was designed around three goals motivated by prior work:

\paragraph{Connect process to location in the text.}
Version histories typically present document states 
chronologically, requiring writers to scan entire versions to find specific changes. Inspired by Hill et al.'s edit wear \cite{hill92edit}, which visualizes accumulated interaction directly on the artifact, we designed \textit{revision 
highlighting} to link version history states to their direct locations in the text. Areas of revision are highlighted on top of the text based on the frequency of edits, and highlights can be clicked to filter the version history to show only the edits made at that location. This makes revision content and intensity visible in-situ, connecting areas of work directly to their locations in the text, and provides writers access to the history of a specific passage without navigating the full timeline.

Highlighting areas of revision not only visualizes effort spent on each phrase and sentence but also suggests areas that invite future attention and effort. This design embodies Schön's theory of professional work, the \textit{reflective conversation with work materials} \cite{schon83reflective}. In his framing of reflection-in-action, Schön describes how interfaces convey and encourage certain problem-settings \cite{schon83reflective}. Edit wear highlighting enables problem-setting through \textit{instability and tension} \cite{hill92edit}. A deeply colored area might represent ``struggle with word choice,'' while an unhighlighted area might denote ``a lack of attention.'' Both shape future actions that the writer might take. 

\paragraph{Focus on relevant states within a verbose process history.}
A single writing session can produce over a hundred process states \cite{guo20effects}, making the version history difficult to parse sequentially. Inspired by Archive
of Our Own’s filter-by-tags navigation \cite{fiesler16archive}, we designed \textit{process filtering} to let writers constrain the history by writing process type (e.g., viewing only planning or revision states) or by narrative entity (e.g., a character name), drawing on the Flower and Hayes cognitive model \cite{flower81cognitive} for process categories and named-entity recognition \cite{honnibal20spacy} for entity categories. This makes it feasible to navigate long histories at a meaningful scope and level of detail. %

\paragraph{Enable reflective inquiry over process.}
\textit{Search} lets writers pose reflective, plain-language questions over the full revision history and receive answers that summarize and link back to relevant process states
in the form of “citations” (``[ ]'').
Queries can target process (\textit{``What part of the writing was I stuck on the longest?"}) and content (\textit{``When did I change the tone of the introduction?''}).
This format of asking questions about the writing corresponds to common reflective practices in composition pedagogy, where students are asked to reflect on the processes and meaning of their writing \cite{yancey14writing, yancey16rhetoric}. Posing questions probes the \textit{context}, \textit{exigency}, \textit{process}, and \textit{meaning} of the writing. This sort of reflection represents a means of fostering metacognition \cite{yancey16rhetoric}, which is central to understanding the text and how to make it better. Natural language interaction helps align writing traces to the writer's own conceptual understanding of their writing, such as ``character development,'' ``wording choices,'' and ``points of struggle,'' thereby connecting traces to their practice. Responses were generated using \texttt{gpt-5-mini}, which was chosen for speed over \texttt{gpt-5}, then the frontier model. From pilot testing, we found \texttt{gpt-5-mini} to be sufficient in quality for the constrained writing sessions of the formative study, and importantly, faster than \texttt{gpt-5}, a requirement for the live system walkthrough.

\subsection{Methods}
\label{sec:initialmethods}
 
\textbf{Participants:}
Participants who self-identified as creative writers with at least three years of experience were recruited from institutional communication channels, word-of-mouth, and social media. Fourteen participants were enrolled (9 women, 5 men; mean age $\approx$ 26, range: 21--33). Each session lasted approximately ninety minutes. Participants were compensated \$15 USD per hour. We received IRB approval from our institution.

\textbf{Procedure:}
Participants were asked to write a short story in the system. Three sample writing prompts were provided, along with a link to additional prompts\footnote{\url{https://blog.reedsy.com/creative-writing-prompts/}}; participants could also write on a topic of their own choosing. We asked participants to try to produce a ``final'' draft, as if submitting a manuscript to an editorial, so that we could observe all parts of the writing process, from brainstorming to revision.
After the writing session, the interviewer introduced each feature of the system and had the participants explore their writing history. For \textit{Search}, sample writing reflection questions were provided that participants could choose from, or ask their own questions. Participants used each feature and answered questions (in a semi-structured interview format) related to their usage, interpretation, and assessment of each feature. The interview guide is provided in the appendix (Sec. \ref{sec:formative_supplement}).

\textbf{Analysis:}
We conducted a reflexive thematic analysis \cite{braun06TA} on the interview transcripts. The first and second authors independently coded the first five interviews, compared codes, discussed interpretations, and established a shared understanding of the emerging themes. The first author continued coding the remaining interviews, iteratively grouping codes into larger themes. These themes were discussed and refined by both coders. The initial five-interview round was done to ensure both coders developed a shared understanding before the first author proceeded with the full set.
 
\subsection{Findings}
\label{sec:initialfindings}

The formative study allowed us to explore how writers used the system to understand their writing process, and to identify design considerations for iterating on the system. We organize our findings by the three main interactions: revision highlighting, process filtering, and natural language search. We report the main themes for each and the design considerations that emerged.

\subsubsection{Revision Highlighting Connected Text to Local Process Histories}

Participants conceptualized highlighting as a form of \textit{local version control}\footnote{We did not describe highlighting as version control or histories during the study.} that enabled inspection of prior versions at specific text locations. As P3 noted, it is \sqinline{P3}{kind of like a version control, but [on] a sentence and word level}. Clicking a highlighted segment filtered the version history sidebar to show only states affecting that location, which participants used to check wording decisions (e.g., P4 discovering they had changed ``The snack'' to ``My snack''), reflect on discarded alternatives (e.g., \sqinline{P10}{'foolishly' or 'deeply'? 'Is it a thin layer of ice, or a thin layer of snow?'}), and trace how passages developed over time (e.g., seeing that a passage \sqinline{P13}{ended up in a totally different direction}).

Participants contrasted this in-situ access with the friction of document-level version histories. P6 described the conventional workflow of retrieving a single prior edit from Google Docs: \squote{P6}{I went back into versions, and then I was sort of scouring through what I wanted to... I didn't want to revert to that version, because that would mean it would revert everything up to that point. I just wanted that local change... So, I had to, sort of, put myself into that version, copy and paste that local change... And then go back to the current version.} With highlighting, this reduced to clicking a segment: \sqinline{P6}{I can literally just go to this highlighted text}.

Beyond retrieval, participants read the \textit{depth} and \textit{distribution} of highlights as cues to their writing behaviors. Darker regions indicated effort---places where they \sqinline{P2}{don't feel like I could quite get it right the first time, or the first couple of times}. Clusters at the beginning suggested going back and forth between drafting later sections and revising earlier ones: \sqinline{P13}{I see a lot... at the beginning. I had added a lot of information in the story, and then as I was, like, going back and editing, I was, like, removing a lot of that information}. Highlights at paragraph boundaries signaled restructuring; small highlights signaled wording changes. Whether about struggle (\sqinline{P1}{these are the parts that I struggled [with]}), indecision (\sqinline{P10}{where I, you know, was deciding between things}), or tension around word choice, highlighting connected the visualization to writers' knowledge of their own decision-making, making them confirm \textit{what} concerns they had and \textit{where}.

These interpretations led to \textit{problem-setting} \cite{hill92edit}. Deeply highlighted regions raised questions about whether passages \sqinline{P7}{still need tweaking}, while unhighlighted regions signaled neglect: \sqinline{P7}{I haven't changed some things at all, and they also need tweaking}, prompting P13 to attend to \sqinline{P13}{what I actually haven't edited}.

\paragraph{Design considerations.} Highlighting was effective as a bridge between the text and the version history: participants used it to identify \textit{where} revision had occurred, then clicked to inspect \textit{what} had changed. However, inspection required navigating to the version history sidebar, and the value participants described was overwhelmingly in the quick, in-situ reading of highlight patterns. P4's reaction to seeing a highlight was immediate---\sqinline{P4}{so the purple is concentrated here. What did I revise here?}---suggesting that the moment of curiosity happens at the text, not in the sidebar. This indicated an opportunity to surface local version history closer to the point of interest. Highlighting shows decision-making, but decisions only make sense in the context where they took place.

\subsubsection{Process Filtering Surfaced Hidden Behaviors; Entity Filtering Was Limited}

Planning and pause states were 
particularly interesting because they produce no visible text changes. P10 found it \sqinline{P10}{interesting [that it] captures where I pause, because I don't notice that myself. It's very passive, and it's good something can bring it out}. Filtering by local revision confirmed that participants were \sqinline{P8}{frequently changing or switching words to better frame my story}. Process-type filtering was generally regarded as more useful than entity filtering, showing where each writing process corresponded to in the text.

Entity-based filtering helped some participants trace how characters evolved (\sqinline{P13}{one of the characters or something... had changed overall}) and check consistency. However, its utility was severely limited. In stories centered on a single character, filtering scarcely filtered at all: \sqinline{P2}{David is the protagonist, so, like, everything in the story can be tagged to him}. The on-device NER model produced errors---\sqinline{P3}{I don't get why this is BS, this is PhD, though. Like, what... Is this trying to... guess... the organization and objects?}. %
Most participants felt longer works with more characters and settings would better reveal the benefits of entity filtering [P3, P4, P5, P7, P9, P10, P11, P12, P13].

\paragraph{Design considerations.}
Two findings guided the next iteration. First, the contrast between process filtering and entity filtering suggested that the tagging mechanism needed to handle semantic context and avoid misclassifications. Second, participants' enthusiasm for process filtering pointed toward a broader opportunity: if writers valued filtering by abstract categories like ``planning'' and ``global revision,'' they might also value filtering by other meaningful categories that the current system could not express.

\subsubsection{Natural Language Search Provided Process Insight Grounded in Evidence}

All participants felt \textit{Search} surfaced novel insights about their writing process. During active composition, writers focused on \textit{what} to write rather than \textit{how}, leaving room for \textit{Search} to reveal unnoticed patterns after the fact: \sqinline{P11}{I'm curious about my writing process... because 
when I was writing, I will not pay attention to, okay, how much time here, how much time there was spent}. P8 realized \sqinline{P8}{I have this kind of, back and forth, like changes, when I was drafting the story}. Participants often began with vague intuitions that \textit{Search} confirmed and crystallized; thoughts became \sqinline{P11}{tangible}, and P5 noted: \sqinline{P5}{I had a rough idea, but I wasn't exactly sure what the answer was. But after reading it, I definitely agree}. For P9, the experience went beyond writing to \textit{self-concept}: \sqinline{P9}{it is that point of reaffirming my own thought process... a very metaphysical thing of reaffirming who I am to myself}.

Linking \textit{Search} responses to clickable version history states helped participants connect system output to concrete writing 
actions. This grounding increased perceived trustworthiness: \sqinline{P5}{it's providing you with evidence as to why a certain suggestion or certain comment was made. And, you know, you can look at it too, and you can make a decision for yourself}. Citations also jogged memory---P6, reading a response, recalled: \sqinline{P6}{'The first recorded switch from drafting to revising occurs at Snapshot 4...' Okay, I do remember this specifically. That was a typo}. P9 valued the specificity: \sqinline{P9}{wording-wise, I love the fact that it's picking up the exact words I messed up, or I reused, or I edited out ASAP}. Given awareness that the tool relied on an LLM, this evidence-grounding mattered for trust: \sqinline{P10}{I know it's not... hopefully it's not hallucinating}.

The \textit{Search} interface also stimulated iterative reflection. P7 described a dialogic quality: \sqinline{P7}{it also forces you to ask questions. And to get a better answer, you have to keep getting more specific... It forces you to think a little more clearly about what you're trying to do}. Even when responses were inaccurate, evaluation itself provoked reflection: \sqinline{P7}{I don't know if that's true... But you're right; it is, like, getting me to think about specific moments, and specific language, which I think is good}.

\paragraph{Design considerations.} Three patterns informed the next iteration. First, citations to version states were central to trust and comprehension, but participants had to click citation numbers and locate the corresponding state in the version history sidebar to read its contents. P6's immediate recognition upon reading a cited snapshot---\sqinline{P6}{I do remember this specifically}---and P9's appreciation for seeing \sqinline{P9}{the exact words I messed up} suggested that the cited material itself, not just the reference to it, was what made responses useful. This pointed toward surfacing the content of cited states inline rather than requiring navigation. 
Second, participants valued detailed process summaries, suggesting that longer, more comprehensive analysis could be more useful. Third, P7's observation that getting better answers required one to \sqinline{P7}{keep getting more specific with your questions} indicated that the single-query format constrained the longitudinal part of writing reflection. Writers wanted to follow up, clarify, and go deeper.

\section{Design of \system{}}
\label{sec:IRIS}

\begin{figure}[ht]
\centering
\includegraphics[width=\linewidth]{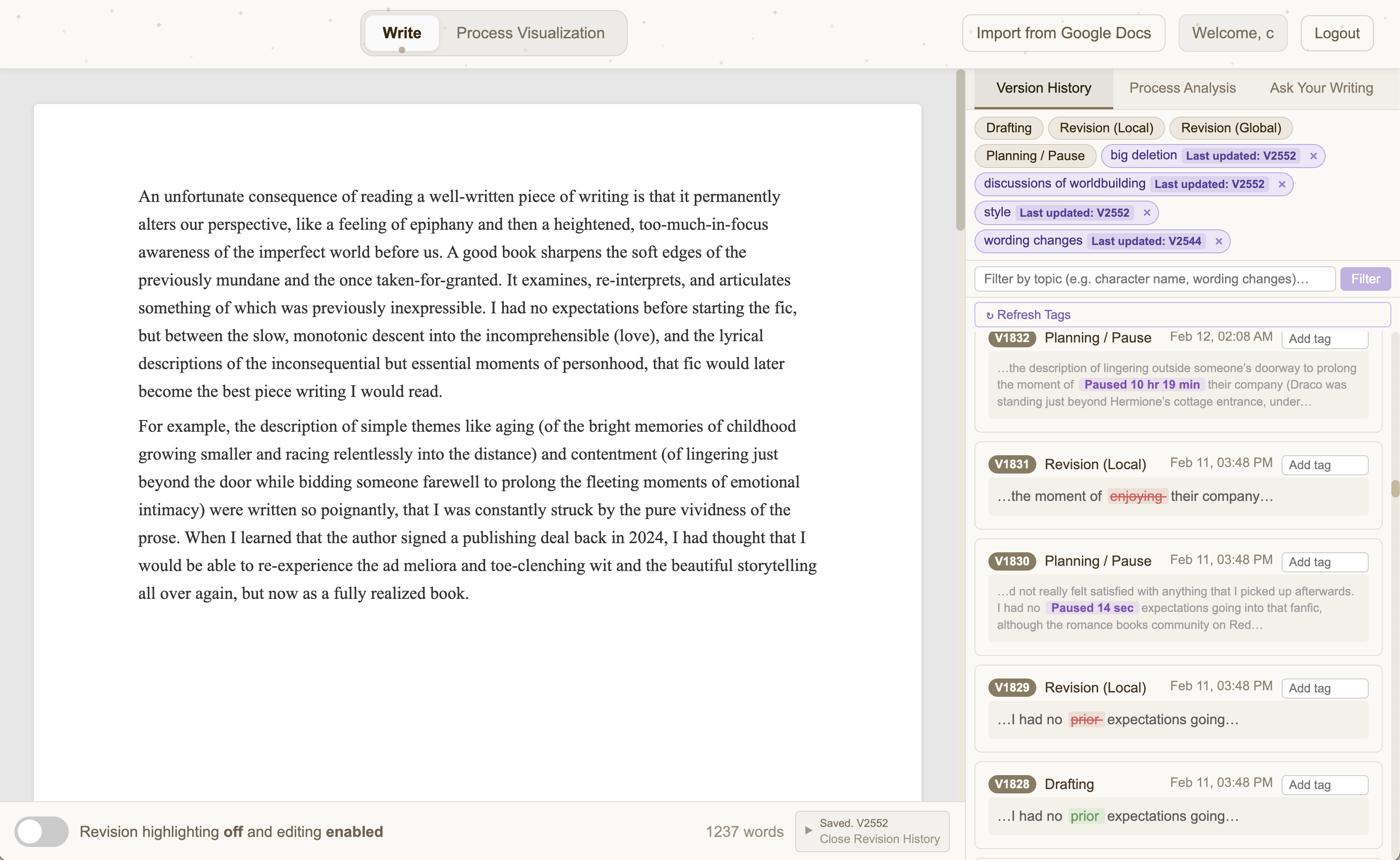}
\caption{\system{} system interface with the side panel opened.}
\label{fig:interface}
\end{figure}

Based on the formative study findings, we refined the prototype into IRIS\footnote{\underline{I}n-situ \underline{R}evision \underline{I}nquiry \underline{S}ystem. A floral name was selected to reference the \underline{Flower} and Hayes Cognitive Process theory \cite{flower81cognitive}. The accent color is purple to mirror an iris.} (Fig. \ref{fig:interface}), a text editor that parses writing processes into version history states from keystroke logs. The design goals and theoretical grounding are described in Section~\ref{sec:initialstudy}; here we describe the implementation of each feature and the changes made in response to the formative findings (Sec.~\ref{sec:initialfindings}).
The individual techniques \system{} draws upon, such as keystroke logging, filtering, and retrieval-augmented generation, are not by themselves new. Rather, its contribution lies in composing these techniques to address new interaction challenges (cf.\ Fogarty's characterization of contribution through composition \cite{fogarty17code}).
First, \textit{aligning traces to the artifact}: process trace analysis has been primarily a post-hoc research tool (Sec.~\ref{sec:rw_capture}), not yet designed for end users. \system{} tracks editing indices during writing, anchoring each change to its location in the current text. Second, \textit{aligning traces to the writer's own understanding}: the output should relate to the writer's conceptual understanding of their writing to make process data situated and interrogable.

\subsection{Inferring Process from Keystroke Logs}

To couple visualizations to writing behaviors, we require a real-time, non-intrusive signal about what the writer is doing. We implement this by using the process states from the Cognitive Process Model \cite{flower81cognitive} with parsing heuristics from Zhang et al. \cite{zhang21using}.
Keyboard events and pauses provide a useful signal without the need for specialized equipment or software, such as eye-tracking hardware, or manual action from the user, which can be disruptive to flow \cite{csikszentmihalyi00beyond}. %

Zhang et al.'s taxonomy of writing states 
\cite{zhang21using} operationalizes the cognitive model into observable states by segmenting keystroke logs using text production metrics, pause behavior, and editing actions. High pause durations preceding a burst of text correspond to \textit{planning}; rapid text production with minimal pausing corresponds to \textit{translating}; and, editing-heavy segments (demarcated by more than three delete actions) correspond to \textit{reviewing}, which further divides into \textit{local revision} (edits near the current insertion point) and \textit{global revision} (edits at distant locations or involving punctuation) \cite{zhang21using}. We adapted the algorithm for streamed (live) input, %
with hyperparameters following prior implementations \cite{guo20effects, zhang21using}. This parsing is unchanged from the formative study prototype.

\subsection{Revision Highlighting}
\begin{figure}[ht]
\centering
\includegraphics[width=\linewidth]{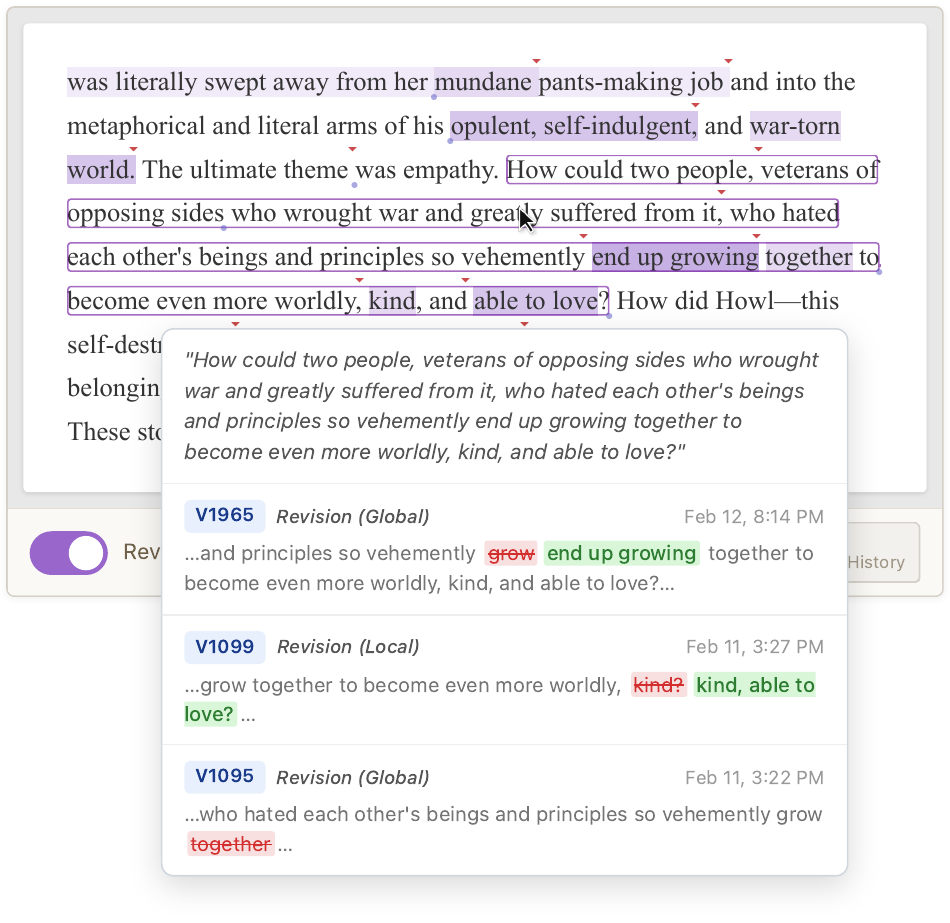}
\caption{Highlight interface in sentence mode, with a tooltip showing the progression of the hovered sentence.
}
\label{fig:highlighting}
\end{figure}

Revision highlighting overlays revision intensity directly on 
the text by progressively shading segments based on the 
frequency of edits at each location (Fig. \ref{fig:highlighting}). 
This overlay allows writers to identify areas of stability and tension \cite{hill92edit}. A deeply 
colored region indicates sustained revision, while an 
unhighlighted region indicates text that has not been revisited. Highlight opacity scales linearly with edit count, capped at four overlapping highlights for readability.

In the formative study, participants valued the in-situ connection between text and process (Sec.~\ref{sec:initialfindings}) but found it disruptive to rely on the sidebar for inspecting specific version states. Participants' reactions to highlights were immediate and localized---curiosity arose at the text, not in the sidebar. To address this, we redesigned the interaction so that hovering over any highlighted segment displays a tooltip showing the local version history at that location. Each entry in the tooltip shows the version number, the inferred process type, and an inline diff displaying what was removed and added. This design further connects the writing process to the notion of \textit{wear}---peeking at what changed---by hovering the wear itself. Additionally, red triangles above the text and purple dots below the text correspond to revisions with no additions (i.e., deletions) and pause states; both can be hovered.
The tooltip can display version entries either per-version (one entry per version state affecting the hovered location) or grouped by sentence.
\textit{Cut} and \textit{paste} was implemented as a delete and insert, which meant that version histories would not travel with moved text, although the ``cut'' text would be shown with a hoverable red deleted marker.

\subsection{Process Filtering}
\begin{figure}[h]
\centering
\includegraphics[width=\linewidth]{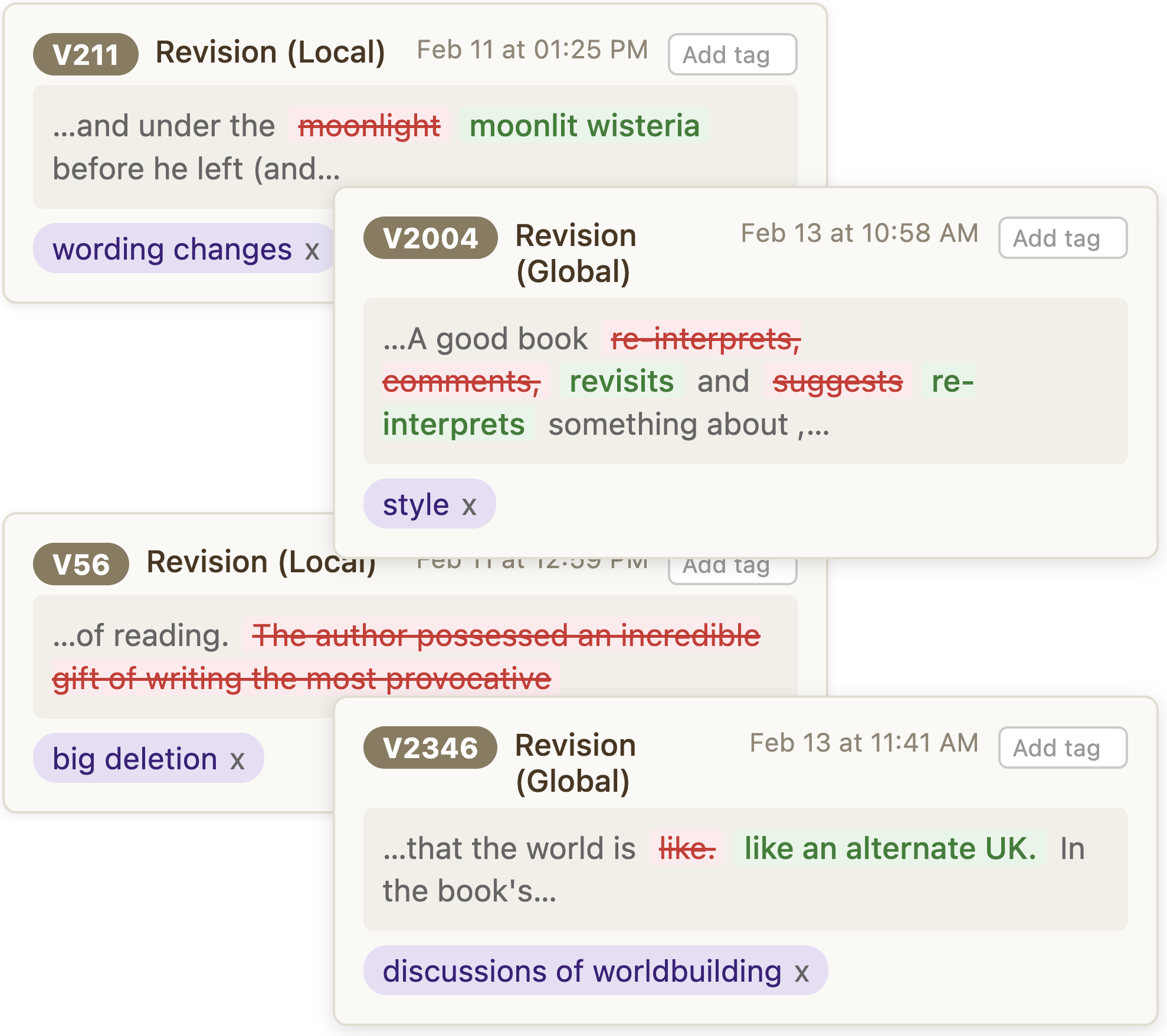}
\caption{Examples of filtered version states. 
The purple tags below are user-defined filters assigned by the system.
}
\label{fig:filtering}
\end{figure}
Writers can filter the version history by \textit{process type} using toggle chips at the top of the sidebar (Fig. \ref{fig:teaser}B). These correspond directly to the inferred process categories. Filtering by process type was well-received in the formative study, particularly for surfacing planning and pause states that produce no visible text changes.

In contrast, entity-based filtering in the formative prototype (Fig. \ref{fig:design_changes}, top) used on-device named-entity recognition, which produced frequent misclassifications and could not handle semantic context. For instance, it attributed a line of dialogue to a mentioned character, not the speaker (P14). We replaced the NER-based tagger with an LLM-based approach. Writers type a conceptual query (e.g., ``wording changes,'' ``character development,'' a character name, etc.) into the filter field. The system sends the query along with a batch of version state diffs to \texttt{gpt-5.2}. Tagged states appear with a removable chip in the sidebar; toggling the chip then filters the history. Tags can be refreshed as new version states are recorded and can be manually added and deleted for each state.

This approach supports abstract, semantically meaningful filters that the rule-based system could not express (Fig. \ref{fig:filtering}). ``Wording changes'' finds revisions where one word was replaced by another; ``tone'' finds changes to attitude and emotion.

\subsection{Natural Language Inquiry}

The formative study found that natural language search provided metacognitive insight grounded in process evidence, but three tensions emerged (Sec.~\ref{sec:initialfindings}). Participants valued cited version states but found navigating to them disruptive, wanted additional detail from the response, and felt that the single-query format constrained the naturally iterative character of process reflection. We addressed these by splitting the functionality into two complementary interfaces: \emph{Process Analysis} for detailed, topic-based summaries, and \emph{Ask Your Writing} for conversational inquiry.

\subsubsection{Process Analysis}

Process Analysis (Fig. \ref{fig:teaser}C) lets writers add topics or questions, such as ``wording changes,'' ``style,'' or ``where did I get stuck?'', and receive a detailed, LLM-generated summary of the writing process pertaining to that topic. The system sends the full version history and the topic to \texttt{gpt-5.2}, which returns the structured  analysis.

The key design changes from the formative prototype are the response content and how citations are presented. In the prototype, responses were brief, and citations were clickable reference numbers that scrolled the version history sidebar to the corresponding state. Participants had to leave the response to verify claims. In \system{}, the response is longer and more detailed, and weaves cited version states within the text as inline expandable groups. Citations can be hovered to show a list of corresponding writing states. The LLM was also instructed to directly quote the content diff in its response. This meant the evidence for a claim was visible alongside the claim, without requiring navigation.
Topics can be re-analyzed, producing an updated summary that reflects new writing. New version states may be relevant to past queries.

\subsubsection{Ask Your Writing}

Ask Your Writing (Fig. \ref{fig:teaser}D) uses the same pipeline as Process Analysis but provides a 
chat interface where writers pose questions and receive shorter, more focused responses with the same citation format. 
Unlike Process Analysis, which produces a comprehensive summary, Ask Your Writing is a chat interface that supports follow-up questions, clarification, and iterative deepening---the dialogic interaction pattern that P7 described but the single-query prototype could not support. The conversational format is also designed to be quicker to parse because its responses are much shorter and interspersed with user-input.

\subsection{Process Visualization}
\begin{figure}[h]
\centering
\includegraphics[width=\linewidth]{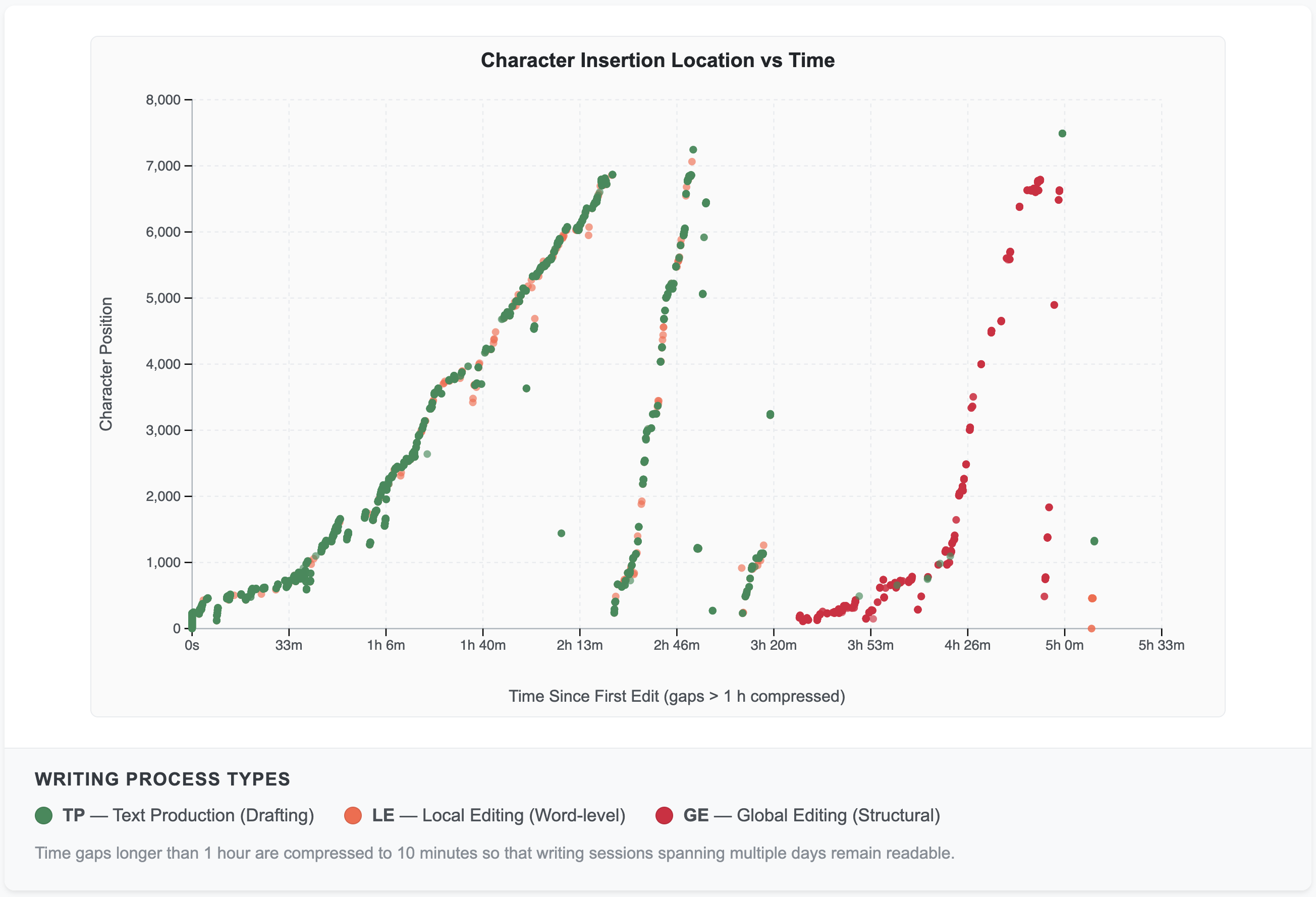}
\caption{Text insertion location visualization. Planning\slash pause states are hidden because they do not modify the text.}
\label{fig:visualization}
\end{figure}

To evaluate the writing process algorithm's \cite{zhang21using} alignment with authorial writing processes as well as writer interpretation, \system{} includes a text insertion graph (Fig. \ref{fig:visualization}), similar to Itero \cite{turkay18itero} but with distinct colors to differentiate writing process type. As this visualization provides a global view of the writing, coupled with \textit{what the writer was doing}, we wanted to see how writers were interpreting broader strokes of their writing process. Participant examples can be found in the appendix (Fig. \ref{fig:process-viz-grid}).

\subsection{System Deployment}
Our tool is implemented as a web application. The frontend was implemented with React, alongside a backend Flask server that handles all data, including user login, version history state management, text analytics, and OpenAI queries and processing.

\section{Methods}
To evaluate how writers use our revised system and to understand how they use process-informed version histories in longer-term writing, we conducted a diary study with creative writers who used the system over the course of one to three weeks.

\textbf{Participants: }
Participants who self-identified as creative writers were recruited from institutional communication channels, word-of-mouth, and social media. Thirteen participants were enrolled; two dropped out due to a time conflict, resulting in 11 participants included in the analysis (6 women, 5 men; mean age $\approx$ 28, range: 22--36). Participants were compensated \$15 USD per hour during the diary study.

\textbf{Procedure: }
Unlike the formative study, where we asked participants to imagine they were submitting a short creative piece for an editorial, we encouraged participants to write something typical for their own writing practice, such as a chapter of a longer story. 

After the pre-interview, in which the interviewers introduced the study and the system, participants were given up to three weeks to draft their writing. During this time, participants completed between four and seven writing sessions (74 total sessions), each targeting thirty minutes. Diary length was set by session count, not a fixed period. Spacing varied with writers' schedules and writing habits, hence the range. Participants could check web resources (e.g., a thesaurus), but were asked not to use external LLMs, in order to focus the conversation on our tool. At the conclusion of the diary component, we performed hour-long semi-structured exit interviews, grounded in the diary entries and focusing on how the tool was used in their daily writing practice. The interview guide and diary questions are provided in the appendix (Sec. \ref{sec:diary_supplement}). 

\textbf{Analysis: }
We performed an inductive thematic analysis \cite{braun06TA} on the interview transcripts. Two authors independently coded six interviews, then discussed their codes to share interpretations and reach consensus. The first author then open-coded the remaining interviews and iteratively grouped the codes and quotes into potential themes. A third author coded three interviews and participated in ongoing discussions of theme construction. The themes were discussed and refined again by all coders.

\section{Results}

Writers used \system{} to support classic creative versioning behaviors, like returning to old ideas
and having the security to make ``bold'' choices (W5). They also leveraged it for familiar AI paradigms, like getting unstuck 
or asking for feedback.
Echoing results from the formative study (Sec. \ref{sec:initialstudy}), participants also used \system{} to reflect on their decision making and gain insight into their writing behaviors. In this section, we focus on 
how the interface's presentation of trace data enabled interpretation of the writer's writing process.

Final texts averaged 1,952 words (min: 568; max: 5,390), with the system recording an average of 879 version states per writer (min: 333; max: 1,912). 
Participants averaged 7.9 Process Analysis queries (min: 3; max: 14) and 3.5 Ask Your Writing chat sessions (min: 1; max: 9). Of the chat sessions, 52\% involved more than one conversational turn, averaging 10.2 turns (min: 1; max: 45).
Writing topics and a Markov transition table of sequential process states are provided in the appendix (Sec. \ref{sec:appendix_markov}, \ref{sec:appendix_topics}).

\subsection{Relating Trace Data to Writing Process}

\system{} presents writing trace data directly to the writer in three forms: a graph of keystroke locations (Fig. \ref{fig:visualization}); highlights in the text (Fig. \ref{fig:highlighting}); and citations of past versions (Fig. \ref{fig:teaser}D). A key design decision in the revised interface of \system{} is a focus on connecting representations of traces to the text itself by adding the version content to all highlights and citations. %
Participants related trace data to prior advice, experiences, and self-perceptions of how they write. %

For example, multiple participants used the Process Visualization graph to reassess their self-perceptions of how they typically write in light of past advice, experiences, or aspirations.
W11 connected the insertion graph to past advice: \sqinline{W11}{A lot of my writing instructors in the past have been like, [name] just write, like, don't edit, don't correct your sentences... Don't try to do the same thing simultaneously. It's a bad habit}. The graph shows that: \squote{W11}{
There's proof here that I don't just write. I sit and edit words, and I do a lot of tinkering as I'm writing... 
So this was actually really useful for me to see and reaffirm that I'm not editing while I write so much as just testing out words or testing the water for how to start a next sentence.
}  W10 and W2 had similar reflections, seeing phases of drafting and editing that \sqinline{W10}{I have pretty much trained myself to do} and \sqinline{W2}{that's always been a part of my process}. The design revision linked quotes directly from all four interface views, facilitating writers' engagement with their own past text. This direct connection to the changes \sqinline{W2}{made me feel pretty grounded}. W3 described the citations in Process Analysis as \sqinline{W3}{it gave again a really good breakdown with specific examples...telling you exactly where it's getting its information from}.

Constructing interpretations of the system responses was an important step in relating traces 
to the author's understanding of their writing. For instance, Process Analysis cited both remaining and deleted text, which supported \textit{recall} of past iterations: \sqinline{W5}{It cites things that I deleted, and so I, remember the things that I deleted pretty well, actually, because of that}. W5 interpreted the presence of insertions and deletions as \sqinline{W5}{it giv[ing] me information about the things I liked and disliked}.
The knowledge that the system could recognize exploration of alternatives was encouraging (W5, W6): \sqinline{W5}{It made me more bold in my...
writing decisions}.

Interpreting the system responses involved \textit{reflection}---thinking about past writing and goals. As W6 mentioned, Process Analysis \sqinline{W6}{help[ed] me reflect on what I wrote, and how, if I can like, was that version better, or was this version better, or stuff like that}. Reflection also extended to the writing process itself. Queries to Process Analysis encouraged \sqinline{W3}{me [to] wonder, how much of this (planning/pausing) is necessary and how much is procrastination?}.
When reflecting at the end of the study, W6 described both Process Analysis and Ask Your Writing as helping with the \sqinline{W6}{writing flow} due to the system's ability to aid in process and writing recall.
Generally, we observed more frequent usage of the system's features towards later writing sessions, during which writers tended to shift towards revision due to having more text written.

\subsection{Confirming Interpretations of Writing Process}
\label{sec:results_confirming}

Writers frequently turned to the system to confirm their interpretations or beliefs about their writing. These instances of confirmation from the system then shaped writers' composition decisions and self-understanding.   
For example, when W2 consulted the system on character development (\sqinline{W2}{Was she likable? Was she interesting? Like, what are her characteristics that I was kind of discovering as I was writing?}), they interpreted the answer as confirmation that their goal had been achieved and decided to move on: \sqinline{W2}{(Process Analysis) gave me an idea that the character was developing in the way I wanted her to. So... that wasn't something I would have to focus on as much}.

Confirmation of one's writing processes occurred when the system articulated what a writer was doing in terms they had not used themselves. W5 viewed themselves as an unconfident writer: \sqinline{W5}{there's always, like, that voice in my head, and that's like, this doesn't sound too good. I don't really like this, but I'll keep going}. When using Ask Your Writing to discuss the organization of the writing, \sqinline{W5}{I felt like the system put that into better words than I could express myself... It's just a feeling of, like, not liking it. But for the system, it's, like, a very specific thing that I could improve on}. 
For W2, who had concerns with how a character was interpreted, an explanation of how W2 developed that character gave them confidence that the character was being expressed as intended: \squote{W2}{It kind of walked through the different revisions here to show how Madeline was changing as a character. It was helpful for me to see that's like, okay, yes, she is growing into this different type of a person. %
}

The system's retrieval-augmented approach foregrounded the writer's own text in responses. For example, when W6 asked for framing feedback as ``validation of what I think,'' \system{} provided multiple examples from the text: \sqinline{W6}{The system give[s] me like almost 5 to 6 lines [from my history]... It's not just one. You have these, these, these, these, these}. For W6, the system had provided past %
iterations underlying its assessment of essay framing, which W6 then used to reorganize their writing. 
Similarly, the version history grounded the feedback for W3 and felt aligned with what they ``should have thought'': \squote{W3}{some of the suggestions it gave for adding things in, like with specific sources, citations, and why it was important to support those claims... I felt like I should have thought of this, but I didn't.}

The feeling of confirmation then defined future work. When W6 was using Process Analysis to reflect on tone, the system confirmed the writer's intended shift: ``Early snapshots carry a raw, diaristic, self-directed tone with slang... By the later consolidated rewrite, the tone shifts to reflective, essay-like, and more generalized.'' W6 had sought this self-confirmation: \sqinline{W6}{Wait, I need that... I needed it to tell me that, oh, I was changing the tone of it}; the response then shaped their word choice in future writing: \sqinline{W6}{I remembered that I shouldn't use these things (words) that it pointed out}. W4 was partially aware of their overuse of punctuation, which the system confirmed and clarified: \sqinline{W4}{It's not just the fact that I'm using a lot of em dashes, but it's also the frequency, that the proximity of the way they're being introduced... It did make me reflect a little bit, and also help not come up with the same situation in the future}.

\subsection{Challenging Beliefs about Writing Process}

In other cases, writers challenged what they received. Sometimes writers faced the contrasting information as simply an alternative perspective and used that difference to reflect on their writing.
For example, W8 used contrasts between interpretations to think about when the writing might not be working: \sqinline{W8}{I was trying to see if the system interpreted me correctly...am I doing the right job as conveying it to the system or would a human pick it up?}.

As the system would attempt to narrate the writer's actions, sometimes its interpretation overstated what the writer was doing. During an analysis about character writing, the system described it as ``The writer was solving two problems at once, making the sci-fi concept legible and giving Madeline a coherent internal logic that explains why this flier matters to her.'' In response, W2 said: \sqinline{W2}{I mean, like, I guess, yes, technically, that is what I was doing, but that's not how I was thinking about it. I was like, I need to set up why she's going to the birthday party, and that's it}. In this case, the misalignment was encouraging:
\squote{W2}{I think it can be encouraging in a sense that, maybe I'm a better writer than I thought I was. If it thinks I'm doing these kind of more sophisticated thinking techniques and processes as I'm writing, that I don't even notice myself doing... 
It's like, ``Oh, actually, no, you're doing good. You're, like, doing things that authors do!''} Such feedback is important to receive as a writer, as other participants have also emphasized when showing their writing to others.

Tension with the interface surfaced when writers challenged the process analysis tools on a more fundamental level. For instance, even when W1 felt like they had received an accurate depiction of how they write, self-confirmation was \textit{frustrating}: \sqinline{W1}{I have written a lot in the past... I've had a lot of revisions from previously human editors in the past, with similar feedback, so this is something I feel like I already know}. Since they were in an exploratory phase of writing and aware of their writing's flaws, they dismissed the system response 
as an unnecessary call-out to metacognitive knowledge they already had.

\section{Discussion}
Our findings show that writing trace data is useful not only for history-keeping and constructing post-hoc analyses, but also for creating an interpretive space where writers relate their activity traces to their writing beliefs, feelings, and aspirations. We discuss the design implications for process-centered CSTs (Sec. \ref{sec:discussion-process-centered}), alternate representations of process (Sec. \ref{sec:discussion_representations}), and the limitations of this work (Sec. \ref{sec:discussion_limitations}).

\subsection{Process-Centered Creativity Support Tools}
\label{sec:discussion-process-centered}
Writing tools expose history chronologically (e.g., linear version lists in Google Docs) or optimize primarily for collaboration and accountability (e.g., tracked changes in Microsoft Word). \system{} instead centers \textit{process}---how text was produced and considered---as the principal index for navigation. 
This reframing points to several design directions that extend prior work, while also suggesting a broader class of process-focused creativity support tools.

\paragraph{A non-generative role for AI in creative work.} 
The dominant paradigm for AI is direct access to a possible final product. In writing, this includes systems that draft, complete, brainstorm, and rephrase text. Although productive for many, this paradigm potentially displaces writers' own ideation for evaluation \cite{bhat26reactive} or unproductively amplifies a writer's own thoughts---W3 describes their experience with a popular chat LLM as dealing with \sqinline{W3}{marketing copy}.

Instead of providing outputs, a process-centered CST such as \system{} foregrounds metacognitive benefits. Articulation and discovery of one's writing process are \sqinline{W2}{motivating and encouraging} and offer \sqinline{W5}{a lot more room for thinking} than tools \sqinline{W5}{that just give you what you want}. Feelings of self-confirmation can productively move a writer towards \textit{creative momentum}: ``creative struggle without setbacks in execution'' \cite[p.~2]{zhou23creative}.

\paragraph{Towards process-responsive tools.}
Writing processes are idiosyncratic, and tools that treat all writing practices similarly will result in misalignment at times. This suggests an opportunity for \textit{process-responsive} tools that adapt not only to what has been produced, but also how it was produced. More broadly, this reframes what it means for a CST to be ``intelligent.'' Current intelligent writing assistants are intelligent about \textit{language} \cite{lee24design}. In contrast, a process-responsive tool would be intelligent about \textit{practice}---understanding not just the text but also the trajectory of decisions, attention, and effort that produced it. This is a difference between a tool that can revise a paragraph and a tool that knows the writer has already revised that paragraph thirty-nine times\footnote{Ernest Hemingway rewrote the last page of \textit{A Farewell to Arms} thirty-nine times \cite{plimpton58ernest}.}. We see this as a productive frame for creativity support research: designing tools that are responsive to processes of creative work, rather than only to the artifacts that work produces.

\paragraph{From artifact versions to a palette of prior decisions.}
Creative practitioners use past versions as a ``palette of materials'' not just for reverting, but also for exploration and reflection across project lifecycles \cite{sterman22towards}. Version control can then become a place for material interaction and exploration of past work \cite{rawn23understanding}. 
Here, we emphasize the role of \textit{interpreting creative decisions and processes} as an essential part of the creative work itself. Being able to see one's earlier writing revisions as \sqinline{W11}{testing out words} not only relates trace data to personal writing processes, but also updates the writer's understanding of their own actions. 
Murray suggests that all writing is autobiographical; writing reflects personal experience but also \textit{reinterprets} it: ``We become what we write. That is one of the great magics of writing'' \cite[p.~71]{murray91all}. Enabling new ways to understand how writing unfolds then centers creativity support on the reflexive process of creativity. Process traces function as material for ``reflective conversation'' \cite{schon83reflective} that can be used to interrogate and understand one's practice.

Hammad et al. recently formalized \textit{creative activity traces} as records of creator data produced over the course of a creative process \cite{hammad26tracing}. \system{} is an instantiation of this concept for writing and 
suggests three design principles for creativity support tools that involve trace data in their design and data pipeline.

\textit{Bind traces to the text and to the writer's understanding.} Process data is most useful when it is anchored both to the artifact (where in the text did it happen?) and to the writer's conceptual frame (what was I trying to do?). Revision highlighting and inline citations point to locations; filtering and analysis in writerly vocabulary (e.g., shifts in tone) point to the writer's understanding. Without the first, traces remain as abstract, verbose logs; without the second, traces remain as difficult-to-interpret metrics.

\textit{Make claims verifiable and contestable.} Users could inspect the cited version states to judge the claim against the data and their own experience. This contestability supported knowledge construction through the gap between the system interpretation and felt process (Sec.~\ref{sec:discussion_representations}). Tools that present process interpretations should therefore make evidence accessible and inspectable, so that confirmation and challenge remain available to the writer.

\textit{Use writerly concepts as the navigational index, not only temporal or metric-based aggregation.} Conventional version histories organize states by time and temporal proximity, whereas \system{}'s filtering and natural language querying instead let writers navigate history through their own conceptual vocabulary (e.g., a character's name, where they got stuck, etc.). This reframing allowed users to move to specific versions much more quickly, and also ask abstract questions like ``how did my introduction evolve,'' connecting version history to reflective practices valued by composition pedagogy \cite{yancey16rhetoric}.

\subsection{Representations of Writing Process}
\label{sec:discussion_representations}

\system{} represents the writing process via the Flower and Hayes cognitive process model \cite{flower81cognitive}, using Zhang et al.'s keystroke parsing heuristics \cite{zhang21using} to infer planning, translating, and reviewing states. This framing enabled the system's core interactions (linking version history states to cognition), but our studies also reveal the boundaries of this representation and the broader space of process models that future tools might draw upon.

\paragraph{The gap between traces and felt process.}
Keystroke logs are proxies for cognition, and participants identified moments where process inference diverged from their experience. In the formative study, P6 contrasted the ``absolute'' time spent revising with their own perception of effort spent towards revision. P9 characterized this as the ``measurement gap'' between thought process and what the system could infer. W2's reaction to the ``solving two problems at once'' characterization similarly revealed a gap between personal experience and system understanding.

Yet, these gaps do not invalidate the approach. Participants often found the system's interpretation useful despite the mismatch, but this gap highlights what keystroke-level traces \cite{zhang21using} and the cognitive process model \cite{flower81cognitive} cannot capture: intentions, rhetorical goals, and the social and situational dimensions that post-process \cite{trimbur94taking, kent93paralogic} and genre \cite{bazerman94systems} theorists suggest.

\paragraph{Normative effects of making process visible.}
Any tool that represents process also shapes its interpretation. Li et al. describe this as the \textit{normative ground} of creativity support tools: 
``the way a tool structures its practitioners’ ideas, goals, and intentions, shaping how someone thinks, acts, and expresses themselves'' \cite{li23beyond}. 
In our studies, this effect surfaced as the \sqinline{P9}{specific mold} of the interface that \sqinline{P10}{seems to try and want to put me in a box more than meet me where I'm at}.
Because we took a process-centric approach to writing, users had to work in the ``walled gardens'' \cite{li23beyond} of the tool's handling of writing process. This normative ground is not unique to our system; it is a general element of creativity support tools \cite{li23beyond}. But, it underscores the importance of positioning for such tools (such as \system{}) as one perspective among many, rather than as authoritative accounts of how writing should work.

Other frameworks would foreground different aspects of process. A tool grounded in Larsen-Ledet et al.'s concept of \textit{text functions} \cite{larsenledet20collaborative} would distinguish between ``material'' and ``interim work'' (there are eight defined text functions) rather than ``planning'' and ``translating'' \cite{flower81cognitive}. Fuzzy Linkography \cite{smith25fuzzy} suggests that interaction logs can be structured into a design-centered representation of the creative process and summarize phases of the creative work.

Our findings also suggest a distinction between \textit{procedural} and \textit{semantic} evolution. \system{} foregrounds procedural evolution (when and how the writer drafted, revised, and paused) because these states are directly observable from keystrokes. But W2's use of the system to track how a character became more likable (Sec. \ref{sec:results_confirming}) points to the concept of semantic evolution---how the \textit{ideas} in the text evolved. Semantic changes are entangled with procedural traces (as text revisions contain meaning and carry motives), but surfacing them would require interpretation beyond what keystroke parsing could provide, such as distinguishing a wording substitution from an intentional shift in characterization. Future tools might support this through improved semantic differencing over revision content, or by inviting writers to annotate why a revision was done, going beyond mechanical changes within the text.

\subsection{Limitations and Future Work}
\label{sec:discussion_limitations}

In the formative study, participants wrote for approximately half an hour, and in the diary study, participants wrote for approximately 3.5 hours. While these studies allowed us to collect a broad set of live writing data and feedback, they may not be sufficiently long to capture the entire lifecycle of writing, particularly for long-form writing. Future work might investigate how process-focused tools scale to entire project lifetimes, where reflecting on past versions can reveal changes in motivation, interests, and craft \cite{sterman22towards}.

Making writing traces visible, particularly through a natural language interface, introduced judgments of writing process that led to feelings of self-consciousness for two participants. W5 wrote more slowly, and W1 felt the presence of an ``observer'' that made the writing feel relatively performative. These effects raise design questions about when and how creative processes might be displayed to the user---how to provide transparency and useful information while minimizing the effects of being observed. W4's practice of toggling revision highlighting only during periods of revision suggests that writer-controlled visibility may help. 

Future work might investigate the effects of specific features using ablation studies and suggest minimally invasive features. \system{} used a commercial LLM, which raises privacy and ethics concerns, such as unauthorized training on user data with limited recourse for intellectual property infringement \cite{gero25creative}. A self-hosted model would help protect user privacy and intellectual property. Our local-model prototyping fell short on response quality, speed, and required context size, which were needed for our diary study, which involved much larger texts and trace data, to run smoothly.

Besides exploring longer periods of practice, future work might support collaborative writing (e.g., sharing and group-reflecting on process slices, combining multiple writers’ traces; W9 speculated that seeing ``the tit-for-tat'' between coauthors would reveal editing dynamics otherwise not shown in coarse version diffs); writing pedagogy (e.g., how to encourage reflective writing among students); and unanticipated usage (e.g., how writers reappropriate parts of the system and integrate them into their own workflows \cite{li23beyond}).

\section{Conclusion}

Writing tools shape not only text but also the writing process and writing metacognition. By surfacing the writing process, \system{} invites writers to recall, inspect, and reinterpret their own writing processes. 
We found that writers treated process traces as interpretive material, reading activity records in light of personal experience and aspirations. Confirmation of nascent process knowledge provided both validation of the writing and encouragement to the writer. Challenges from the system prompted writers to reflect on how their writing was coming across. In both cases, the value of presenting writing traces was derived from the connection to the writer's own text and writing goals.
Designing with process at the center opens a generative space for creativity support tools that align with the multitude of ways that meaning can be expressed. At the same time, we caution that process inferences carry normative assumptions that designers must acknowledge.

The reflective engagement of this work was inspired by the practices of great writers like Leo Tolstoy and Roald Dahl, who rewrote and reflected on their writing extensively, not just to refine language but to reshape the very goals of their work \cite{murray73maker}. In her novel \textit{Orlando}, Virginia Woolf penned: ``Every secret of a writer's soul, every experience of his life, every quality of his mind is written large in his works'' \cite{woolf95orlando}. The reflective nature of composition, then, is not merely a byproduct of writing, but central to creative expression. %

\begin{acks}

We are grateful to the members of the PICL lab, particularly James Eschrich, William Goss, and Hannah Kim, as well as Michael Kang, for their helpful insights, discussions, and support.

\end{acks}
\bibliographystyle{ACM-Reference-Format}
\bibliography{sample-base}

\appendix
\onecolumn
\section{Supplementary Results}
\subsection{Process Type Transition Probabilities}
\label{sec:appendix_markov}

 \begin{table}[h]
  \centering
  
  \begin{subtable}{\linewidth}
  \centering
  \begin{tabular}{l cccc}
  \toprule
   & \multicolumn{4}{c}{To} \\
  \cmidrule(lr){2-5}
  From & Planning & Drafting & Local Rev. & Global Rev. \\
  \midrule
  Planning     & 0.00 & 0.67  & 0.27  & 0.07  \\
  Drafting     & 0.74 & 0.00  & 0.22  & 0.04  \\
  Local Rev.   & 1.00 & 0.00  & 0.00  & 0.00  \\
  Global Rev.  & 1.00 & 0.00  & 0.00  & 0.00  \\
  \bottomrule
  \end{tabular}
  \caption{Formative study}
  \label{tab:transition-matrix-formative}
  \end{subtable}

  \vspace{8pt}

  \begin{subtable}{\linewidth}
  \centering
  \begin{tabular}{l cccc}
  \toprule
   & \multicolumn{4}{c}{To} \\
  \cmidrule(lr){2-5}
  From & Planning & Drafting & Local Rev. & Global Rev. \\
  \midrule
  Planning     & 0.00 & 0.74  & 0.22  & 0.04  \\
  Drafting     & 0.64 & 0.00  & 0.27  & 0.09  \\
  Local Rev.   & 1.00 & 0.00  & 0.00  & 0.00  \\
  Global Rev.  & 1.00 & 0.00  & 0.00  & 0.00  \\
  \bottomrule
  \end{tabular}
  \caption{Diary study}
  \label{tab:transition-matrix-diary}
  \end{subtable}
  
  \caption{Pooled first-order Markov transition matrices. Cell values are mean participant-normalized transition probabilities.}
  \label{tab:transition-matrix}
  \end{table}
  
Version state transitions are modeled here as a first-order Markov chain over the four writing states (planning, drafting, local revision, and global revision) derived from keystroke logs following the interkey interval (IKI) classification procedure \cite{zhang21using, guo20effects}. A transition is counted between each pair of temporally consecutive version states within the same writer and writing session. \system{} must estimate version state boundaries with only the text produced so far due to input being streamed, does not aggregate existing and committed version states (which would otherwise destroy them), and groups segments immediately following the onset of local or global revision within a burst (which results in local revision and global revision always transitioning to planning/pause, since bursts are always separated by planning/pause), which are deviations from Zhang et al.'s post-hoc algorithm \cite{zhang21using} made to better support live writing with \system{}. %

\subsection{Supplementary Diary Participant Data}

\begin{figure*}[t]
\centering

\begin{subfigure}[b]{0.48\linewidth}
  \centering
  \includegraphics[width=\linewidth]{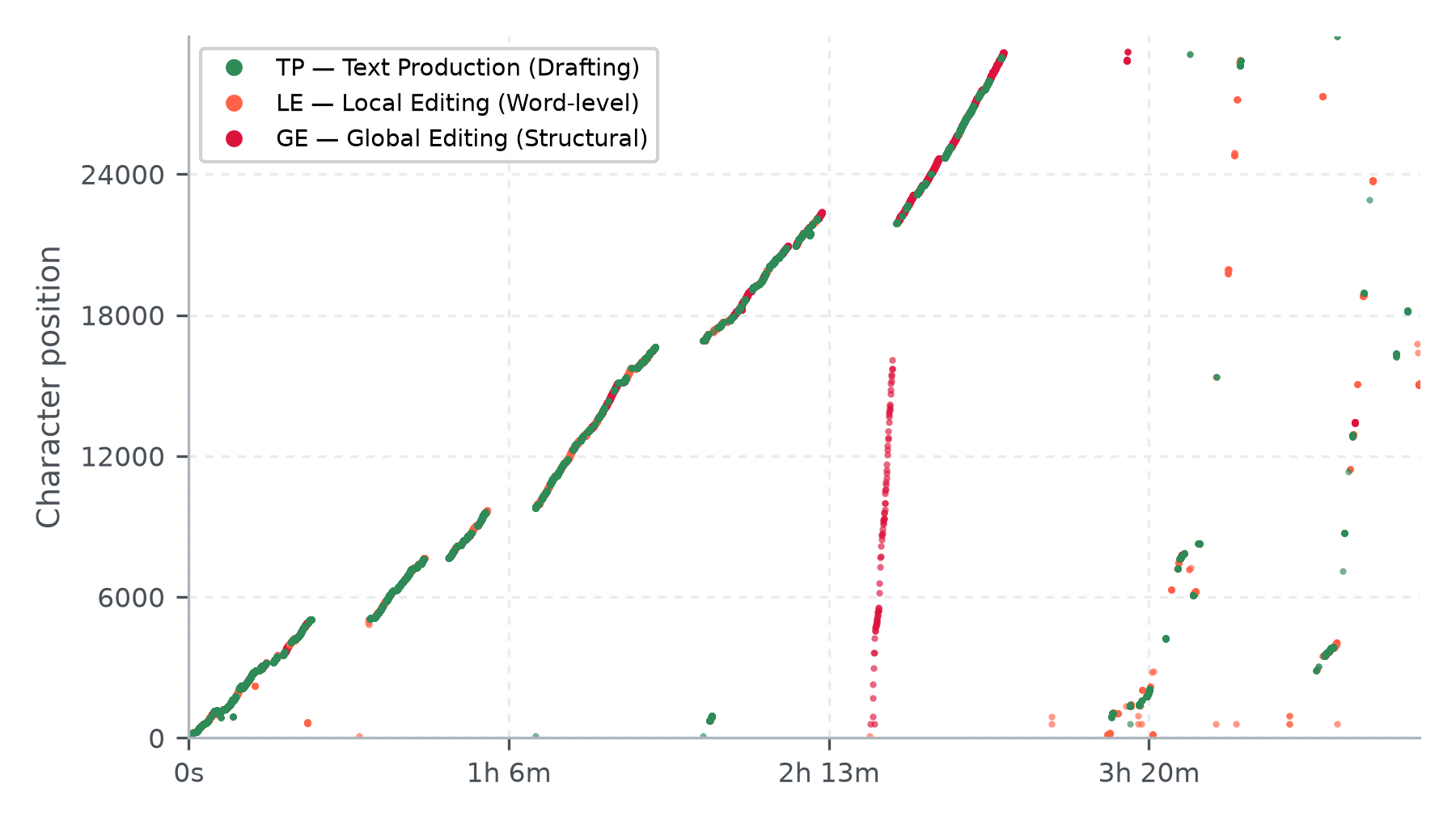}
  \caption{W2}
  \label{fig:process-viz-w2}
\end{subfigure}
\hfill
\begin{subfigure}[b]{0.48\linewidth}
  \centering
  \includegraphics[width=\linewidth]{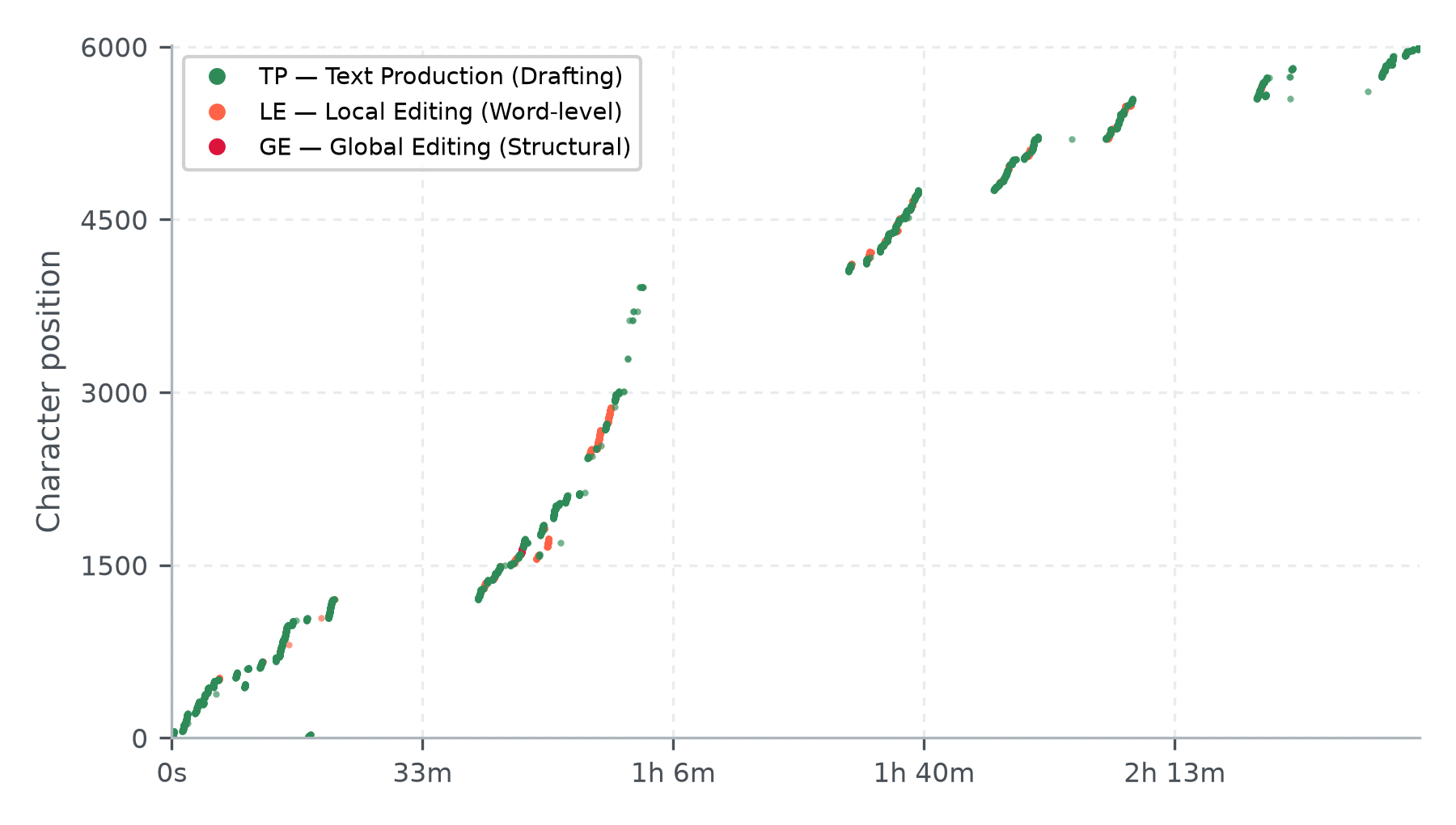}
  \caption{W3}
  \label{fig:process-viz-w3}
\end{subfigure}
\vspace{6pt}

\begin{subfigure}[b]{0.48\linewidth}
  \centering
  \includegraphics[width=\linewidth]{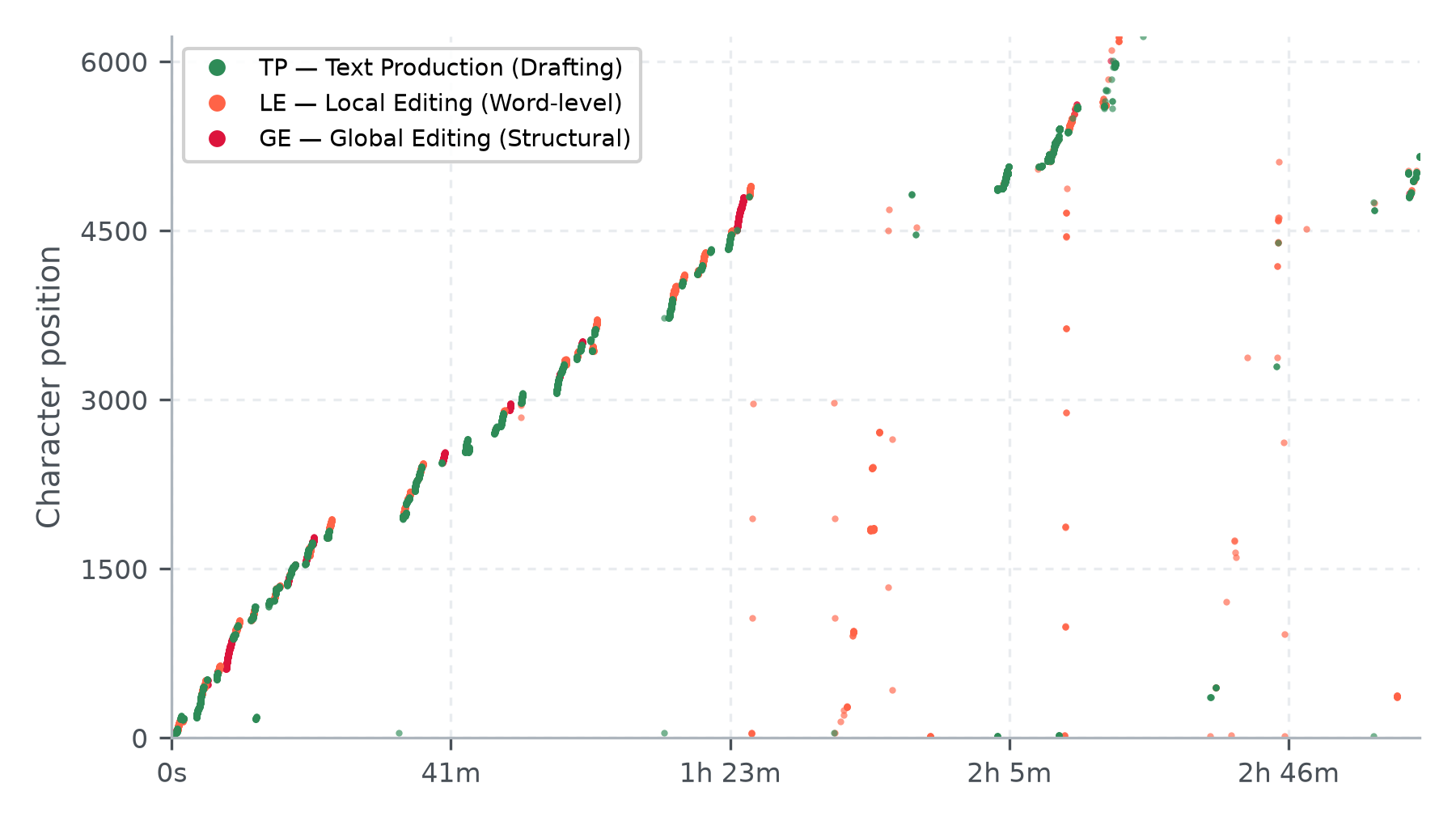}
  \caption{W7}
  \label{fig:process-viz-w1}
\end{subfigure}
\hfill
\begin{subfigure}[b]{0.48\linewidth}
  \centering
  \includegraphics[width=\linewidth]{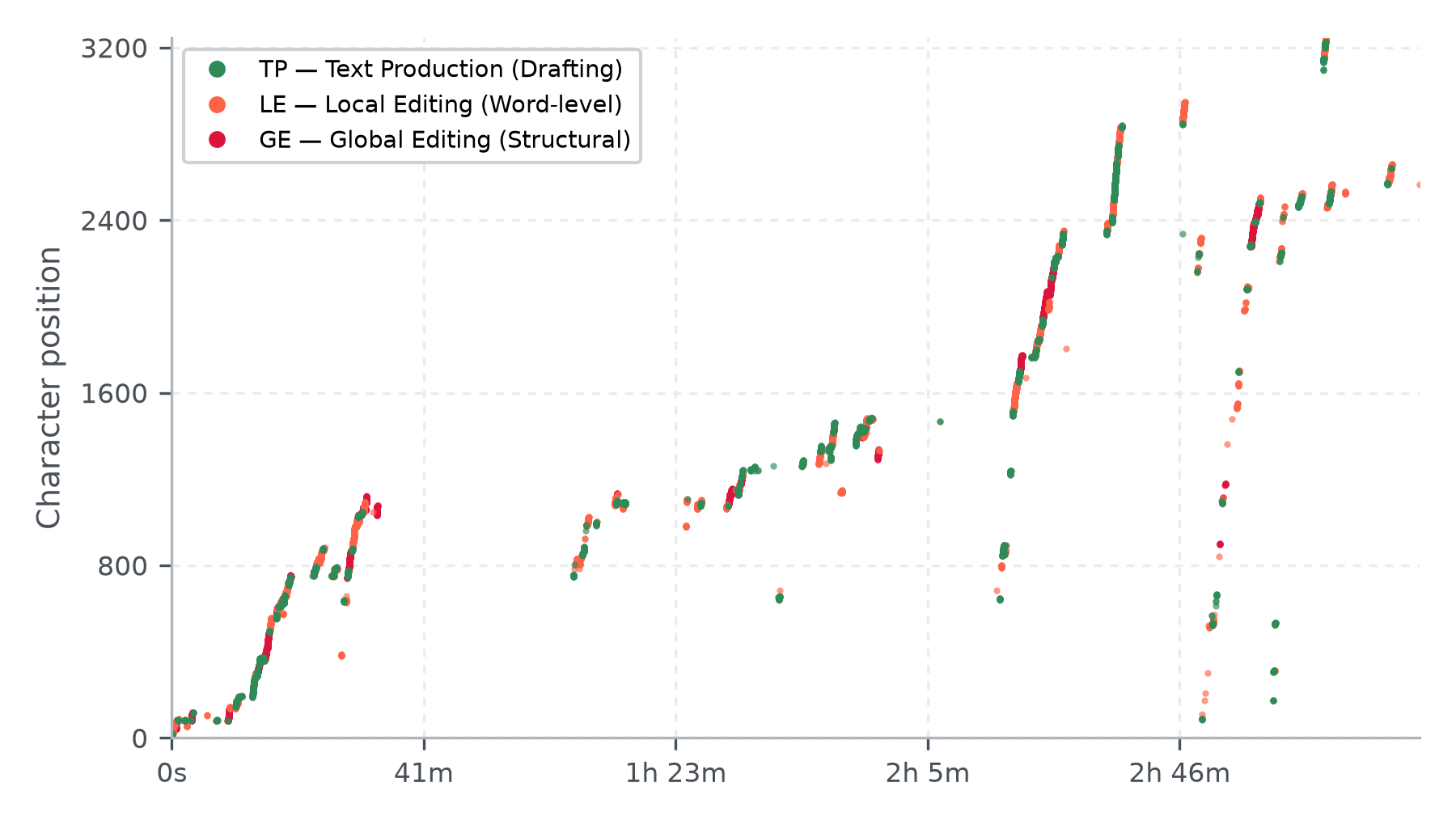}
  \caption{W11}
  \label{fig:process-viz-w4}
\end{subfigure}

\caption{Examples of process graphs from four participants in the diary study. These graphs show character insertion position vs. time, colored by inferred writing process. Gaps longer than one hour are compressed on the x-axis to improve readability, the same format that participants saw in the \system{} interface.}
\label{fig:process-viz-grid}
\end{figure*}

Here we describe the texts written by the participants during the diary study (Table \ref{tab:participants}). We also provide a selection of four representative process visualization graphs from diary study participants (Fig. \ref{fig:process-viz-grid}).

\label{sec:appendix_topics}

\begin{table*}[h]
\centering

\begin{tabular}{l l r r r r}
\toprule
\textbf{ID} & \textbf{Writing Topic} & \textbf{Sessions} & \textbf{Avg. Session (min)}  & \textbf{Num. Version States} & \textbf{Word Count}\\
\midrule
W1  & Reflection on a family vacation & 7 & 25 & 910 & 1014  \\
W2  & Sci-fi short story & 7 & 30  & 948 & 5390 \\
W3  & News article on internet scams & 7 & 25 & 499 & 961 \\
W4  & Fictional short story & 7 & 27 & 1348 & 1256 \\
W5  & Personal reflection on instant ramen & 7 & 27 & 1087 & 860 \\
W6  & Personal reflection on romance & 7 & 29 & 333 & 1383 \\
W7  & Personal reflection on ethnic cuisine & 7 & 28 & 519 & 1104 \\
W8  & Fanfiction of a popular YA series & 7 & 28 & 350 & 1490 \\
W9  & Continuation of a longer story & 7 & 26 & 1082 & 2195 \\
W10 & Fanfiction of a popular YA series & 7 & 40 & 1912 & 5247 \\
W11 & Fictional short story & 4 & 64* & 681 & 568 \\
\bottomrule
\end{tabular}
\caption{Diary study participation. Session lengths overall averaged 30.47 minutes. *W11 wrote a one-hour session and a two-hour session, and their session average would be 37 minutes if averaged across seven hypothetical sessions.}
\label{tab:participants}

\end{table*}

\subsection{Design Evolution}
\begin{figure*}
\centering
\includegraphics[width=\linewidth]{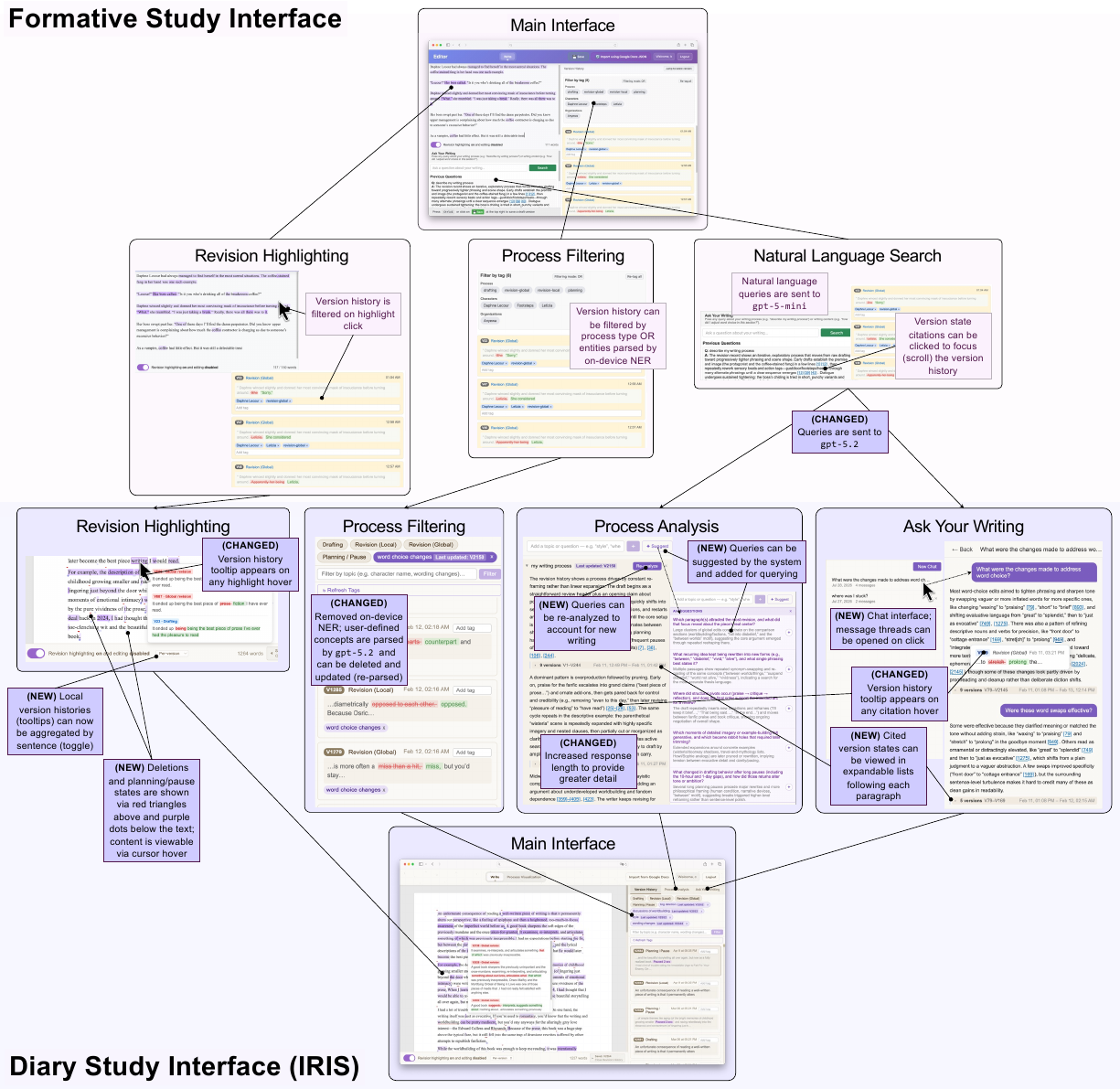}
\caption{Design changes to \system{} following the formative study.}
\label{fig:design_changes}
\end{figure*}

Figure \ref{fig:design_changes} shows key interface changes between the tool used for the formative study and the final design of \system{} used in the diary study.

\lstset{
  basicstyle=\small\ttfamily,
  breaklines=true,
  breakatwhitespace=false,
  frame=single,
  columns=fullflexible,
  keepspaces=true,
  aboveskip=0.5em,
  belowskip=0.5em,
  literate={⏎}{{\ensuremath{\hookleftarrow}}}{1}
}

\section{Formative Study Materials}
\label{sec:formative_supplement}

\subsection{Semi-Structured Interview Guide}

\paragraph{Overall Writing Process}
\begin{itemize}
    \item What was the goal for your story? What were you trying to write?
    \item What was going well?
    \item What was difficult?
    \item What was the overall process you took?
\end{itemize}

\paragraph{Revision Highlighting}
\textbf{[brief walkthrough of edit wear functionality]}\\
\textbf{[participant engages in edit wear]}
\begin{itemize}
    \item What are areas that were highlighted in your story? (What do they mean or show?)
    \item How do you interpret what's being highlighted? Does it connect to your own writing process? If so, how?
    \item Does the highlighting correspond with how you remember your writing?
    \item Do you notice any patterns? Places where highlights are clustered or sparse? Describe.
    \item If highlighting could be enabled elsewhere (e.g.,\ Google Docs, Word), when might you use it? If not, why not?
\end{itemize}

\paragraph{Filtering}
\textbf{[brief walkthrough of filtering functionality]}\\
\textbf{[participant engages in filtering (e.g., clicks on tags)]}
\begin{itemize}
    \item What do the tags represent?
    \item What do the filtered version histories show to you?
    \item Do the tags help narrow down the version history? Does the tagging feel functional?
    \begin{itemize}
        \item (If not) Why don't you think they work in the way you expected?
    \end{itemize}
    \item When do you think you might choose to use the filtering? If not, why not?
\end{itemize}

\paragraph{Natural Language Search}
\textbf{[brief walkthrough of search]}\\
\textbf{[participant is provided list of sample questions]}\\
\textbf{[participant asks at least 3 questions]}

\noindent After each question is asked:
\begin{itemize}
    \item Why did you choose this question?
    \item How did you feel about the response?
    \item What would you use this information for? If not, why not?
\end{itemize}

\noindent After all Search questions have been asked:
\begin{itemize}
    \item How do you feel about the concept of asking questions about your writing process?
    \item What, if any, did you gain from engaging with your writing process like this?
    \item When would you decide to ask questions?
\end{itemize}

\subsection{Sample Writing Reflection Questions for Search}
\paragraph{Time and effort}
\begin{itemize}
    \item Where did I spend the most time revising? \cite{chamberlain18Draftback,laudel14DocuViz}
    \item What parts of the draft had the least attention? \cite{laudel14DocuViz,uncndRevising}
    \item When did my writing pace slow down the most? \cite{chamberlain18Draftback,laudel14DocuViz}
    \item What did I change repeatedly but keep reverting? \cite{chamberlain18Draftback}
\end{itemize}
\paragraph{Revision}
\begin{itemize}
    \item Show me places of heavy revision. \cite{laudel14DocuViz}
    \item Where did I tighten word choice the most? \cite{mccowanndFour}
    \item Find where my voice/tone changed significantly. \cite{mccowanndFour}
    \item Where did I expand vs trim sentences? \cite{mccowanndFour}
    \item Show changes that clarify argument / adjust voice / tighten pacing. \cite{mccowanndFour}
    \item Where did I reorder paragraphs/sections? \cite{mccowanndFour}
\end{itemize}
\paragraph{Stuckness and breakthroughs}
\begin{itemize}
    \item What was I stuck on the longest? \cite{chamberlain18Draftback,uconn20Sample}
    \item Where did I backtrack or undo a change? \cite{chamberlain18Draftback}
    \item Find a breakthrough moment that unblocked me---what changed? \cite{chamberlain18Draftback,uconn20Sample}
    \item When did the central idea/thesis first appear? \cite{uncndRevising, chamberlain18Draftback}
\end{itemize}
\paragraph{Character and setting development}
\begin{itemize}
    \item How did I develop the character [character] over time? \cite{mccowanndFour}
    \item What did I change about [setting/scene]? \cite{mccowanndFour}
\end{itemize}
\paragraph{Planning the next session}
\begin{itemize}
    \item What are good restart points for tomorrow? \cite{uconn20Sample,chamberlain18Draftback}
\end{itemize}
\textit{Sources were not shown to participants. Sources provide the context for the questions. The reflection questions were informed by Nancy Sommers' and Donald Murray's work on revision \cite{sommers80revision,murray73maker}.}

\subsection{Writing Prompts}

\begin{itemize}
    \item A character finds out their parents blog about them. \footnote{\url{https://archive.nytimes.com/learning.blogs.nytimes.com/2011/02/25/would-you-mind-if-your-parents-blogged-about-you/}}
    \item A character obtains a device that tells them exactly what choices to make in order to lead the ``happiest'' life possible. Some of these choices get hard to make. \footnote{\url{https://www.reddit.com/r/WritingPrompts/comments/4hzddg/wp_you_obtain_a_device_that_tells_you_exactly/}}
    \item Your story begins in a taxi, and involves an old enemy and Valentine's Day. \footnote{\url{https://www.creative-writing-now.com/short-story-ideas.html}}
    \item Reedsy Creative Writing Prompts \footnote{\url{https://blog.reedsy.com/creative-writing-prompts/}}
\end{itemize}

\textit{Sources were not shown to participants. Sources provide the context for the writing prompts.}

\section{Diary Study Materials}
\label{sec:diary_supplement}

\subsection{Semi-Structured Interview Guide}

\begin{itemize}
    \item \textbf{How did the participant use the tool? When do they use it? Why?}
    \begin{itemize}
        \item \textbf{Show me a moment when you used a feature\ldots}
        \begin{itemize}
            \item What were you trying to do (learn, seek)?
            \item What interface actions did you take?
            \item When in your writing process did you do that?
            \item What did you find out by doing that?
        \end{itemize}
    \end{itemize}

    \item \textbf{How did using it change the participant's writing?}
    \begin{itemize}
        \item \textbf{Show me a moment when\ldots}
        \begin{itemize}
            \item What was different about your writing process using this tool?
            \item After using a feature, what did you do? Was that influenced by the tool?
            \item What do you think about your writing process now?
            \begin{itemize}
                \item Did you use it to deliberately reflect?
                \item Did it facilitate reflection when you weren't planning to?
            \end{itemize}
            \item Did it impact the writing outcome?
        \end{itemize}
    \end{itemize}

    \item \textbf{Effectiveness of the tool?}
    \begin{itemize}
        \item Did it seem accurate?
        \item Was it helpful?
    \end{itemize}

    \item \textbf{Attitude towards this use of AI?}
    \begin{itemize}
        \item How would you compare this writing tool to other AI writing tools you know about?
        \item Does the distinction (if any) change how you feel about this usage of AI?
    \end{itemize}

    \item \textbf{Ease of using the version history to see past writing}
    \begin{itemize}
        \item Does it change how you feel about going back to previous versions? Does it feel similar or different?
    \end{itemize}
\end{itemize}

\end{document}